\documentclass[aps, prb, twocolumn, superscriptaddress, floatfix]{revtex4-2}

\usepackage{float}
\usepackage{graphicx}
\usepackage{amsmath,amssymb}
\usepackage{tikz}
\usepackage{braket}
\usepackage{pgfplots}
\pgfplotsset{compat=1.17}
\usepackage[version=4]{mhchem}
\usepackage{physics}
\usepackage{subfigure}
\usepackage{subcaption}
\usepackage{enumitem}
\usepackage{mathrsfs}

\usetikzlibrary{
  shapes.geometric,
  shadows,
  patterns,
  perspective,
  decorations,
  decorations.text
  }

\usetikzlibrary{decorations.markings,calc,decorations.pathreplacing}
\usetikzlibrary{shadings,decorations.pathreplacing}

\usepackage{xcolor}

\definecolor{linkcolor}{RGB}{0,0,255}      
\definecolor{citecolor}{RGB}{10,155,55}     
\definecolor{urlcolor}{RGB}{0,0,255}      

\usepackage{hyperref}
\hypersetup{
  colorlinks   = true,
  linkcolor    = linkcolor,
  citecolor    = citecolor,
  urlcolor     = urlcolor,
  linktoc      = all,
  pdfborder    = {0 0 0},
}

\usetikzlibrary{arrows.meta}

\begin{document}

\title{Multichannel Kondo Effect in Superconducting Leads}
\author{Pradip Kattel}
\email{pradip.kattel@unige.ch}
\affiliation{Department of Quantum Matter Physics, University of Geneva, Quai Ernest-Ansermet 24, 1211 Geneva, Switzerland}
\author{Abay Zhakenov}
\author{Natan Andrei}
\affiliation{Department of Physics, Center for Material Theory, Rutgers University, Piscataway, New Jersey 08854, USA
}

\begin{abstract}
The traditional multichannel Kondo effect  takes place  when several gapless metallic electronic channels interact with a localized spin-$S$ impurity, with the number of channels $n$ exceeding the size of the impurity spin, $n>2S$, leading to the emergence of non-Fermi liquid impurity behavior at low temperatures. Here, we show that the effect can be realized even when the electronic degrees of freedom are strongly correlated and gapped. The system under consideration consists of a single spin-$\frac{1}{2}$ impurity coupled isotropically to $n$ spin singlet superconducting channels realized by one-dimensional leads with quasi-long-range superconducting order. The competition between the Kondo and  superconducting fluctuations induces  multiple distinct ground states and boundary phases depending on the relative strengths of the bulk and boundary interactions. Using the Bethe Ansatz technique, we identify four regimes: an overscreened Kondo phase, a zero-mode phase, a Yu–Shiba–Rusinov (YSR) phase, and a local-moment phase with an unscreened impurity, each with its own experimental characteristic. We describe the renormalization-group flow, the excitation spectrum, and the full impurity thermodynamics in each phase. Remarkably, even in the presence of a bulk mass gap, the boundary critical behavior in the Kondo phase is governed by the same exponents as in the gapless theory with the low-energy impurity sector flowing to the $SU(2)_n$ Wess–Zumino–Witten (WZW) fixed point, and the impurity entropy  monotonically decreasing as a function of temperature. In both the overscreened Kondo and zero-mode phases, the residual impurity entropy is $S_{\mathrm{imp}}(T \to 0) = \ln[2\cos(\pi/(n+2))]$. In the YSR and unscreened phases  on the other hand  the impurity entropy exhibits non-monotonic temperature dependence and is effectively free at low temperatures with $S_{\mathrm{imp}}(T \to 0) = \ln 2$.
\end{abstract}

\maketitle

\section{Introduction}

A single quantum impurity can profoundly alter the many-body properties of a host system, giving rise to emergent correlated phenomena. The archetypal Kondo effect~\cite{nozieres1980kondo} is a case in point. Understanding the behavior induced by a localized impurity is of fundamental importance and has been the subject of extensive theoretical and experimental investigation across a wide range of quantum materials and model systems. These include impurities embedded in metallic hosts~\cite{kondo2012physics,hewson1997kondo,andrei1980diagonalization,andrei1984solution,tsvelick1983exact,komijani2020isolating}, where the impurity is dynamically screened via the Kondo effect; superconducting environments, where impurity-induced bound states and pair breaking phenomena emerge~\cite{Yu,Shiba,Rusinov,pasnoori2022rise,wei2025kondo,moca2021kondo,manaparambil2025underscreened,moca2025spectral,pasnoori2022rise,kattel2025thermodynamics,kattel2024overscreened,kattel2025competing}; and impurities embedded in other one-dimensional correlated systems such as Luttinger liquids~\cite{furusaki1994kondo,furusaki2005kondo,frojdh1995kondo,lee1992kondo,rylands2016quantum}, Mott insulators~\cite{pereira2025tunneling} and spin chains~\cite{wang1997exact,frahm1997open,kattel2024kondo,zhakenov2025thermodynamics}. 
Of particular interest is the effect of bulk correlations on the overscreened Kondo effect, where a localized spin is overscreened by multiple channels of conduction electrons, giving rise to non-Fermi liquid behavior and boundary criticality beyond the Landau paradigm~\cite{NozieresBlandin1980,andrei1984solution,tsvelick1984solution,AffleckLudwig1993,tang2025two}. Although overscreening is traditionally studied in gapless metallic systems,  new features emerge when such impurities are embedded in gapped superconducting environments~\cite{Yu,Shiba,Rusinov,pasnoori2022rise,kattel2024overscreened,wei2025kondo}. Recent cold-atom advances enable engineering of local moments~\cite{Amaricci2025Engineering}, multicomponent fermions~\cite{Taie2010,Scazza2014,Pagano2014}, and tunable one-dimensional superfluid order~\cite{nascimbene2013realizing}, opening paths to impurity physics in regimes inaccessible in solids.

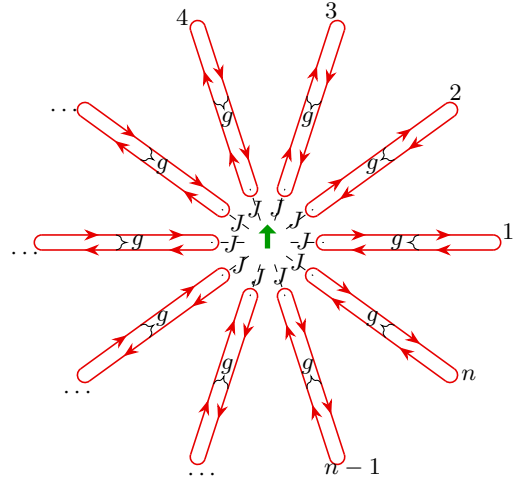
\begin{figure}[t]
\centering
\definecolor{darkred}{rgb}{0.9,0,0}
\begin{tikzpicture}[>=Stealth,scale=1]
  \begin{scope}[xscale=1,yscale=1]
    \def\n{10}
    \def\radius{3}
    \def\gap{0.75}
    \def\offset{0.1}
    \def\wirethickness{0.6pt}

    \node[draw,white,circle,fill=none,minimum size=6mm] (imp) at (0,0) {};

    \draw[line width=2pt,green!60!black] (0,-0.8mm) -- (0,1.5mm); 
    \fill[green!60!black]
  (0,2.4mm) -- (-1.2mm,1.2mm) -- (1.2mm,1.2mm) -- cycle;

    \foreach \i in {1,...,\n} {
      \pgfmathsetmacro{\angle}{360/\n * (\i - 1)}
      \pgfmathsetmacro{\xstart}{\gap * cos(\angle)}
      \pgfmathsetmacro{\ystart}{\gap * sin(\angle)}
      \pgfmathsetmacro{\xend}{\radius * cos(\angle)}
      \pgfmathsetmacro{\yend}{\radius * sin(\angle)}
      \pgfmathsetmacro{\perpx}{-\offset * sin(\angle)}
      \pgfmathsetmacro{\perpy}{\offset * cos(\angle)}
      \coordinate (upStart)  at ({\xstart+\perpx}, {\ystart+\perpy});
      \coordinate (upEnd)    at ({\xend+\perpx},   {\yend+\perpy});
      \coordinate (lowStart) at ({\xstart-\perpx}, {\ystart-\perpy});
      \coordinate (lowEnd)   at ({\xend-\perpx},   {\yend-\perpy});

      \draw[line width=\wirethickness,darkred,
            decoration={markings,
                mark=at position 0.25 with {\arrow{>}},
                mark=at position 0.75 with {\arrow{>}}},
            postaction={decorate}]
            (upStart) -- (upEnd);

      \draw[line width=\wirethickness,darkred,
            decoration={markings,
                mark=at position 0.25 with {\arrow{<}},
                mark=at position 0.75 with {\arrow{<}}},
            postaction={decorate}]
            (lowStart) -- (lowEnd);

      \draw[dashed,black] (imp) -- (\xstart,\ystart);

      \coordinate (gapMid) at ($(imp)!0.5!(\xstart,\ystart)$);
      \pgfmathsetmacro{\labelOffsetX}{-0.3 * sin(\angle)}
      \pgfmathsetmacro{\labelOffsetY}{0.3 * cos(\angle)}
      \node at ($(gapMid) + (\labelOffsetX, \labelOffsetY)$) {\small$J$};

      \pgfmathsetmacro{\midFrac}{0.5}
      \coordinate (midUp)  at ($(upStart)!{\midFrac}!(upEnd)$);
      \coordinate (midLow) at ($(lowStart)!{\midFrac}!(lowEnd)$);
      \draw[decorate,decoration={brace,amplitude=4pt,raise=-4pt}]
        (midLow) -- (midUp);
      \coordinate (braceMid) at ($(midLow)!0.5!(midUp)$);
      \node[inner sep=1pt] at ($(braceMid)!0.15cm!(0,0)$) {\small$g$};

      \pgfmathsetmacro{\dxOuter}{(\xend+\perpx)-(\xend-\perpx)}
      \pgfmathsetmacro{\dyOuter}{(\yend+\perpy)-(\yend-\perpy)}
      \pgfmathsetmacro{\radiusOuter}{sqrt(\dxOuter*\dxOuter + \dyOuter*\dyOuter)/2}
      \pgfmathsetmacro{\angleOuter}{atan2(\dyOuter,\dxOuter)}
      \draw[line width=\wirethickness,darkred]
        (upEnd) arc[start angle=\angleOuter, delta angle=-180, radius=\radiusOuter cm];

      \pgfmathsetmacro{\dxInner}{(\xstart-\perpx)-(\xstart+\perpx)}
      \pgfmathsetmacro{\dyInner}{(\ystart-\perpy)-(\ystart+\perpy)}
      \pgfmathsetmacro{\radiusInner}{sqrt(\dxInner*\dxInner + \dyInner*\dyInner)/2}
      \pgfmathsetmacro{\angleInner}{atan2(\dyInner,\dxInner)}
      \draw[line width=\wirethickness,darkred]
        (lowStart) arc[start angle=\angleInner, delta angle=-180, radius=\radiusInner cm];

      \pgfmathsetmacro{\numRad}{\radius + 0.2}
      \pgfmathsetmacro{\nx}{\numRad*cos(\angle) - 0.15*sin(\angle)}
      \pgfmathsetmacro{\ny}{\numRad*sin(\angle) + 0.15*cos(\angle)}
      \ifnum\i=9
        \node[font=\small] at (\nx,\ny) {$n-1$};
      \else
        \ifnum\i>4
          \ifnum\i<9
            \node[font=\small] at (\nx,\ny) {$\cdots$};
          \else
            \ifnum\i=\n
              \node[font=\small] at (\nx,\ny) {$n$};
            \else
              \node[font=\small] at (\nx,\ny) {\i};
            \fi
          \fi
        \else
          \node[font=\small] at (\nx,\ny) {\i};
        \fi
      \fi
    }
  \end{scope}
\end{tikzpicture}
\caption{Experimentally realizable schematic of $n$ superconducting leads coupled (shown in red) isotropically via the Kondo exchange $J$ to a single spin-$\frac12$ impurity (shown in green). Dashed lines indicate impurity–lead coupling, while braces labeled $g$ denote attractive spin–density-spin-density interactions between electrons of opposite chiralities.}
\label{fig:radial-n-channel-fig-couplings}
\end{figure}

Here we study a spin-$\frac{1}{2}$ impurity coupled via channel-isotropic spin exchange interaction to the edge of $n>1$ one-dimensional gapful superconducting leads. In the absence of bulk superconducting order, the impurity spin is completely quenched by the bulk gapless degrees of freedom and is described by the Fermi-liquid fixed point for $n=1$~\cite{andrei1984solution,tsvelick1984solution}, while for $n>1$, the impurity is overscreened and its low temperatur  behavior is described by an intermediate non-Fermi-liquid characterized by residual fractional entropy~\cite{andrei1984solution,tsvelick1984solution}. However, when superconducting correlations are present the  competition between Kondo screening and superconducting order yields a rich phase diagram exhibiting overscreened Kondo, intermediate zero-mode, Yu–Shiba–Rusinov (YSR), and local moment (unscreened) phases, each with its own experimental signature. We suggest experimentally accessible setups  of non-Fermi-liquid boundary criticality in a gapped bulk, as in Fig.~\ref{fig:radial-n-channel-fig-couplings}. A single impurity in a single superconducting lead has already been realized in various systems~\cite{lee2017scaling,franke2011competition,hatter2015magnetic,yazdani1997probing}.

\section{The Model}
We  consider a single impurity at the edge of $n$ superconducting wires 
described by the $n$-channel chiral Gross-Neveu model~\cite{andrei1980derivation}.
The Hamiltonian of the model takes the form,
\begin{align}
H &= \int_{-L}^0   \mathrm{d}x \bigg[
 -i   \sum_{\substack{r=1,\dots,n \\ \sigma=\uparrow,\downarrow}}
   \big( \psi^\dagger_{r\sigma,+} \partial_x \psi_{r\sigma,+}
        -\psi^\dagger_{r\sigma,-} \partial_x \psi_{r\sigma,-} \big) \notag \\
& + 2g   \sum_{\substack{r,r'=1,\dots,n \\ \sigma,\sigma',\rho,\rho'=\uparrow,\downarrow}}
   \psi^\dagger_{r\sigma,+} (\vec{\sigma})_{\sigma\rho} \psi_{r\rho,+}   
   \psi^\dagger_{r'\sigma',-} (\vec{\sigma})_{\sigma'\rho'} \psi_{r'\rho',-} \notag \\
& + 2J   \sum_{\substack{r=1,\dots,n \\ \sigma,\sigma'=\uparrow,\downarrow}}
   \delta(x)  \psi^\dagger_{r\sigma,-} (\vec{\sigma})_{\sigma\sigma'} 
   \psi_{r\sigma',+} \cdot \vec{S} \bigg].
\label{modelham1}
\end{align}
where we denote the flavor by $r = 1,\dots,n$, spin by $\sigma = \uparrow,\downarrow$, and chirality by $\pm$. The first term in the Hamiltonian is the kinetic energy of the chiral fermions, the second term is an attractive spin-density-spin-density interaction of strength $g$ coupling opposite chiralities in the bulk, and the last term is the boundary Kondo exchange at $x=0$ with coupling $J$ that flips chiralities at the edge and couples to the impurity spin $\vec S$. When $g=0$, the model reduces to the familiar integrable multichannel Kondo problem studied via Bethe Ansatz in Ref.~\cite{andrei1984solution,tsvelick1984solution} and via boundary conformal field theory in Refs.~\cite{ludwig1994exact,parcollet1998overscreened}. Here, we provide an exact Bethe Ansatz solution of the model over all energy scales, establishing the full phase diagram when both the bulk coupling $g$ and the boundary coupling $J$ are non-vanishing, generalizing the single channel solution \cite{pasnoori2022rise}. For a complete picture, we combine the exact solution with perturbative conformal field theory and RG analysis.

\section{Perturbative Analysis}
Before presenting an exact non-perturbative solution of the model using the Bethe Ansatz method, we briefly discuss the perturbative analysis of the model. The spin part of the bulk UV effective Hamiltonian density contains the $SU(2)_{n}$ WZW kinetic term~\cite{affleck1995conformal,AffleckLudwig1993}, proportional in Sugawara form to $\mathcal{J}_{L}^{a}\mathcal{J}_{L}^{a}+\mathcal{J}_{R}^{a}\mathcal{J}_{R}^{a}$~\cite{sugawara1968field}, perturbed in the bulk by the Gross–Neveu term $g \mathcal{J}_{L}^{a}(x)\mathcal{J}_{R}^{a}(x)$ coupling left- and right-moving spin currents with scaling dimension $\Delta_s=2$, and at $x=0$ by the Kondo term $J(\mathcal{J}_L^{a}(0)+\mathcal{J}_R^{a}(0))S_{\mathrm{imp}}^{a}$ coupling the total bulk spin current to the impurity with scaling dimension $\Delta=1$.

Consider the bulk interaction: although gapless in the UV, the coupling $g$ flows to strong coupling via the RG equation $\frac{d g}{d\ln\mu}=-g^2$, since $\mathcal J_{L}^{a}\mathcal J_{R}^{a}$ has scaling dimension $\Delta_S=2$, the beta function vanishes at the tree level whereas its one-loop OPE yields a logarithmic divergence. With $g(\mu_0)=g_0$, the coupling diverges at $\ln(\mu/\mu_0)=-1/g_0$, producing via dimensional transmutation a gap $m=\mu_0 e^{-1/g_0}$ in the IR. The bulk density of states is then $\rho(E)=\frac{E}{\sqrt{E^2-m^2}}$, typical for a gapped superconductor.

For $g=0$, the boundary Kondo operator $J^{a}(\tau)S_{\mathrm{imp}}^{a}$ has scaling dimension $\Delta_S=1$ in 1D, giving a vanishing tree-level beta function. At one loop, the boundary current–spin OPE yields $dJ/d\ln\mu=-J^{2}$, and at two loops, successive insertions add $(n/2)J^{3}$, giving $dJ/d\ln\mu=-J^{2}+(n/2)J^{3}+O(J^{4})$. This predicts an intermediate fixed point $J^{*}=2/n$, valid for $n\gg 1$, describing the overscreened multichannel Kondo model with residual entropy $S_{\mathrm{imp}}=\ln[2\cos(\pi/(n+2))]$. For small $n$, this fixed point is spurious; in particular, for $n=1$ the flow is to the strong-coupling Fermi-liquid limit with the impurity fully screened and $S_{\mathrm{imp}}=0$.

With bulk interaction $g$, the boundary Kondo operator acquires an anomalous dimension, modifying the RG flow. In our cutoff scheme, its UV scaling dimension $\Delta_S=1$ shifts to $\Delta_S=1+g$, yielding $\frac{\pi}{2}\beta_g(J)=-J^2+gJ$, so the bulk term contributes positively and tends to unscreen the impurity\footnote{Note that this has also been calculated perturbatively in Refs.~\cite{affleck1999logarithmic,laflorencie2008kondo}.}. Before presenting the exact solution, we first examine the implications of the boundary coupling one–loop $\beta$–function, 
$\frac{\pi}{2}\beta_g(J) = -J^2 + gJ$. When $J \ll g$, the $\beta$–function is negative, indicating that the coupling flows away from the unstable weak–coupling fixed point. Since $g$ is small compared to the boundary cutoff scale, the renormalization group (RG) flow is expected to approach the same infrared (IR) fixed point as the conventional multichannel Kondo model. In this limit, the impurity becomes overscreened, and the residual impurity entropy assumes the universal fractional value 
$S_{\mathrm{imp}} = \ln[2\cos(\pi/(n+2))]$, characteristic of the overscreened Kondo fixed point described by the $\mathrm{SU}(2)_n$ Wess–Zumino–Witten (WZW) boundary conformal field theory (bCFT) even though the model is gapped in the IR. 

In contrast, when $g\ll J$, the $\beta$–function reverses sign and becomes positive, implying that the effective boundary coupling turns ferromagnetic and hence irrelevant in the RG sense. Consequently, the system flows to a stable local–moment fixed point where the impurity spin remains asymptotically free, with a residual entropy $S_{\mathrm{imp}} = \ln 2$ at zero temperature.

We now present the Bethe Ansatz solution, which allows us to verify the above renormalization-group predictions by solving the model exactly at all energy scales. The analysis reveals that the interplay between the Kondo coupling $J$ and the superconducting pairing strength $g$ gives rise to four distinct boundary phases. For $g \ll J$, where $\beta_g(J) < 0$, the impurity is screened by gapped multiparticle fermionic excitations. For $g \gg J$, where $\beta_g(J) > 0$, bulk superconductivity dominates and the impurity remains unscreened. Remarkably, in the intermediate regime $g \sim J$, two additional boundary phases emerge, characterized by localized bound modes near the impurity—analogous to Yu–Shiba–Rusinov (YSR) states in a BCS superconductor~\cite{Yu,Shiba,Rusinov}.

\section{Exact solution}
We turn to the full non-perturbative solution of the model. Imposing open boundary conditions on an interval $L$, $\psi_{+}(-L) = - \psi_{-}(-L)$ and $\psi_+ (0) = - \psi_- (0)$, we carry out the procedure of   ``dynamical fusion" of flavor strings in the Bethe Ansatz equations. These in turn fuse the field operators into the highest spin composites ~\cite{andrei1984solution, andrei1995fermi,jerez1998solution,PhysRevB.58.7619,zinn1998generalized}. Using this approach, we write down the Bethe Ansatz equations  of Hamiltonian Eq.\eqref{modelham1}: 
\begin{align}
    e^{-2ik_j L}=\prod_{\alpha=1}^M \prod_{\upsilon=\pm} \frac{b+\upsilon\lambda_\alpha+i \frac{n}{2}}{b+\upsilon\lambda_\alpha-i \frac{n}{2}},
\end{align}

where the spin rapidities $\lambda_\alpha $ satisfy  

\begin{align}
&\prod_{\upsilon=\pm}\left(\frac{\lambda_\alpha+\upsilon d+\frac{i}{2}}{\lambda_\alpha+\upsilon d-\frac{i}{2}} \right)\left( \frac{\lambda_\alpha+\upsilon b + i \frac{n}{2}}{\lambda_\alpha+\upsilon b - i \frac{n}{2}}\right)^N\nonumber\\
&=\prod_{\beta\neq \alpha}\frac{\lambda_\alpha-\lambda_\beta+i }{\lambda_\alpha-\lambda_\beta-i }\frac{\lambda_\alpha+\lambda_\beta+i }{\lambda_\alpha+\lambda_\beta-i }.
    \label{maineqn}
\end{align}
 where $M$ dictates the total $z-$component of the given state via $S^z=\frac{N+1}{2}-M$. Having solved the equations and obtained the allowed values of momenta $k_j$ the energy of the eigenstate is given by $E=\sum_{j=1}^N k_j$. As the spectrum is linear, a cutoff $D$ needs to be introduced, requiring $-D \ge k_j$, to render the energy finite. The renormalized theory is then obtained in the scaling limit when $D \to -\infty$ while adjusting the ``running" couplings $g(D), J(D)$ so as to hold fixed all physical quantities such as the spinon mass or the Kondo scale (see below).

\begin{figure*}
    \includegraphics[]{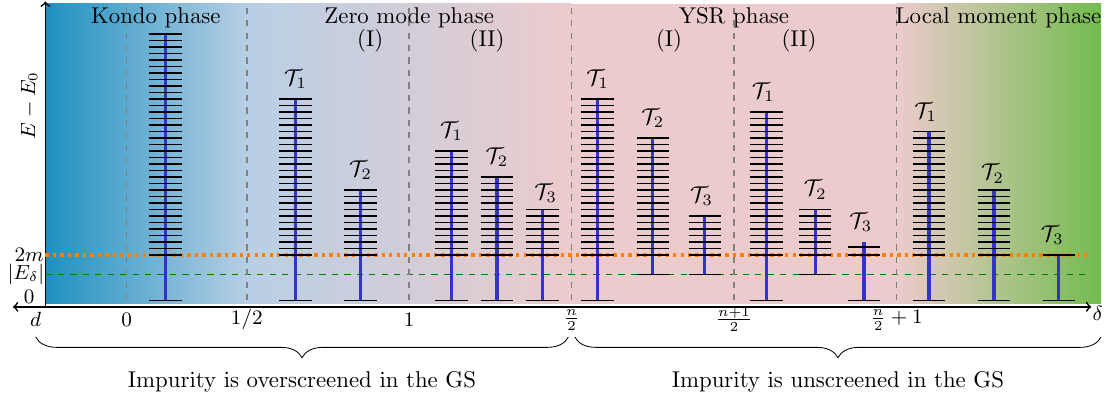}
    \caption{
Phase diagram of the $n$-channel superconducting model with a spin-$\frac{1}{2}$ boundary impurity and even bulk particle number $N$ per flavor. 
The horizontal axis represents the RG-invariant parameter $d(J,g)$, real for $d\ge 0$ and purely imaginary $d=i\delta$ with $\delta>0$ to the right. 
For real $d$ or $0<\delta<\frac{1}{2}$ (Kondo phase), the impurity is overscreened by a multiparticle cloud, and the spectrum consists of a single continuous tower of eigenstates built upon the overscreened ground state. 
For $\frac{1}{2}<\delta<1$ (Zero-mode Phase~I), the emergence of a purely imaginary boundary root $\lambda_d$ splits the spectrum into two distinct towers corresponding to configurations where the boundary mode is unoccupied or occupied. 
For $1<\delta<\frac{n}{2}$ (Zero-mode Phase~II), higher-order boundary strings $\lambda_{\delta,\ell}^{(p)}=\lambda_\delta+i\ell$ with $\ell=1,\dots,p$ and $p=\lfloor\delta+\frac{1}{2}\rfloor$ appear, producing three towers of excitations. 
For $\frac{n}{2}<\delta<\frac{n}{2}+1$ (YSR phase), a quantum phase transition occurs: the impurity becomes unscreened in the ground state, the bound state acquires finite energy, and screening occurs only in the excited sector. 
Three towers persist, but some are lifted in energy due to hybridization with mid-gap states; the nature of this lifting differs between YSR~Phase~I ($\frac{n}{2}<\delta<\frac{n}{2}+\frac{1}{2}$) and YSR~Phase~II ($\frac{n}{2}+\frac{1}{2}<\delta<\frac{n}{2}+1$). 
For $\delta>\frac{n}{2}+1$, screening is impossible at any scale, and the system resides in the local moment phase, where the impurity spin remains asymptotically free and the three towers persist as asymptotically decoupled excitation sectors.
}
\label{fig:phasediag1}
\end{figure*}

In the Bethe Ansatz equations Eq.(\ref{maineqn})
the parameter $d= d(J,g) = \sqrt{b^{2} - 
\frac{2b}{c} - 1} \in \mathbb{R} \cup i \mathbb{R}$  depends on   the bulk  coupling constant $g$   and the boundary coupling constant $J$ through parameterization $      b = \frac{4 - g^{2}}{8 g} $ and $c = \frac{2 J}{1 - \frac{3 J^2}{4}}$.  While both $J$  and $g$ are RG running coupling constants, the parameter $d$ is an RG-invariant parameter which we use below to characterize the various phases of the Hamiltonian. The phase diagram expressed in terms of $d$, is illustrated in Fig.~\ref{fig:phasediag1}. We note that when the bulk parameter $g\to 0$, the Bethe Ansatz equations reduce to those of the multichannel Kondo model studied in Refs.~\cite{andrei1984solution,tsvelick1984solution} and when the boundary parameter $J\to 0$, we recover the Bethe Ansatz equations of the $SU(2)\times SU(f)$ Gross-Neveu model studied in Ref.~\cite{PhysRevB.58.7619}.

 The ground state is formed from length-$n$ string solutions of the Bethe Ansatz equations, each of the form~\cite{andrei1984solution,tsvelick1984solution,andrei1995fermi,jerez1998solution,zinn1998generalized}.

\begin{equation}
\left\{\lambda_{\alpha,j}^{(n)}\right\}_{j=1}^n
=
\left\{\Lambda_\alpha + \frac{i}{2}(n+1 - 2j)
 \middle|  j = 1,\dots,n\right\},
 \label{string-soln}
\end{equation}
with $\Lambda_\alpha \in \mathbb{R}$. Solving the Bethe Ansatz equations, we find that the Kondo coupling flows to an intermediate attractive fixed point, where the impurity is overscreened by the bulk. This conclusion is verified by explicitly computing the impurity thermodynamic quantities and the associated critical exponents, which confirm the non-Fermi-liquid character of this intermediate fixed point.

A complete characterization of the boundary phases is achieved by classifying the  solutions of the Bethe Ansatz equations with fixed  particle number $N$ per flavor so that the total number of particles is $nN$ (we choose $N$ even). Studying the excitations above the ground state we find that a bulk gap opens corresponding to the spinon mass, given by $m=De^{-\pi b} \to D e^{-\pi/2g}$, in the scaling limit. This scale is the same in all phases. On the other hand, the structure of the Bethe Ansatz equations Eq.~\eqref{maineqn} depends on the value of the RG-invariant parameter $d$, and consequently, the boundary phases are determined by its value. As we vary  $d$, the model undergoes an eigenstate phase transition, manifested in the reorganization of the total spectrum into distinct numbers of towers of eigenstates: a single tower in the Kondo phase, two or three in the zero-mode phase, and three in both the Yu–Shiba–Rusinov (YSR) and local-moment phases as shown recently in \cite{zhakenov2025thermodynamics,kattel2025thermodynamics}. By solving the Bethe Ansatz equations across the full range of $d$, we obtain the complete phase diagram and compute the corresponding impurity thermodynamic quantities.

The main findings are as follows. In both the Kondo and zero-mode phases, the impurity is overscreened at low temperatures, and the residual impurity entropy at zero temperature is $S_{\mathrm{imp}}(T \to 0) = \ln[2\cos(\pi/(n+2))]$. In contrast, in the YSR and unscreened phases, the impurity entropy approaches $\ln 2$ in both the zero- and high-temperature limits. In these regimes, intermediate-temperature features arise from boundary string excitations that become thermally active when $T$ is comparable to their characteristic energy $E_\delta$. The impurity entropy varies monotonically with temperature in the Kondo phase, as in the conventional multichannel Kondo effect, while in the other three phases it can display non-monotonic behavior.

This quantitative change in the thermodynamics originates from a phenomenon we refer to as an \emph{eigenstate phase transition}, in which the Hilbert space reorganizes into a different number of excitation towers as the impurity parameter $\delta$ is varied. Each phase can thus be identified by the structure and multiplicity of its eigenstate towers. In the Kondo phase ($0<\delta<\frac{1}{2}$), there is a single continuous tower of excitations built upon the overscreened ground state. In Zero-mode~Phase~I ($\frac{1}{2}<\delta<1$), the emergence of a purely imaginary boundary root $\lambda_d$ splits the spectrum into two towers corresponding to the unoccupied and occupied boundary-mode configurations. In Zero-mode~Phase~II ($1<\delta<\frac{n}{2}$), higher-order boundary strings $\lambda_{\delta,\ell}^{(p)}=\lambda_\delta+i\ell$ with $\ell=1,\dots,p$ and $p=\lfloor\delta+\frac{1}{2}\rfloor$ appear, giving rise to three distinct towers of excitations. Across the transition into the YSR regime ($\frac{n}{2}<\delta<\frac{n}{2}+1$), the impurity becomes unscreened in the ground state while the bound-state excitation acquires a finite subgap energy. The three towers persist but are lifted relative to one another due to the hybridization between the impurity and the mid-gap states. The pattern of this lifting differs between YSR~Phase~I and YSR~Phase~II, reflecting the internal rearrangement of the boundary spectrum. For $\delta>\frac{n}{2}+1$, in the local-moment phase, screening ceases to operate at any scale, and the three towers remain as asymptotically decoupled excitations corresponding to the free impurity spin and the bulk quasiparticles.

\begin{figure*}[t]
  \centering
  \includegraphics[width=0.47\textwidth]{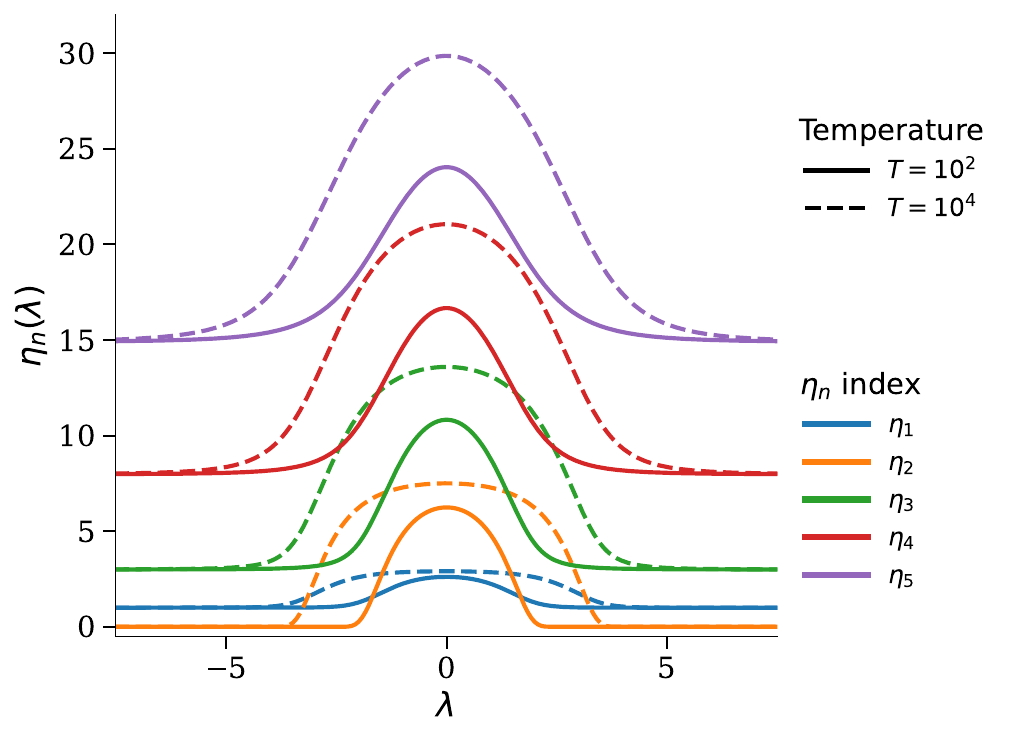}\hfill
  \includegraphics[width=0.47\textwidth]{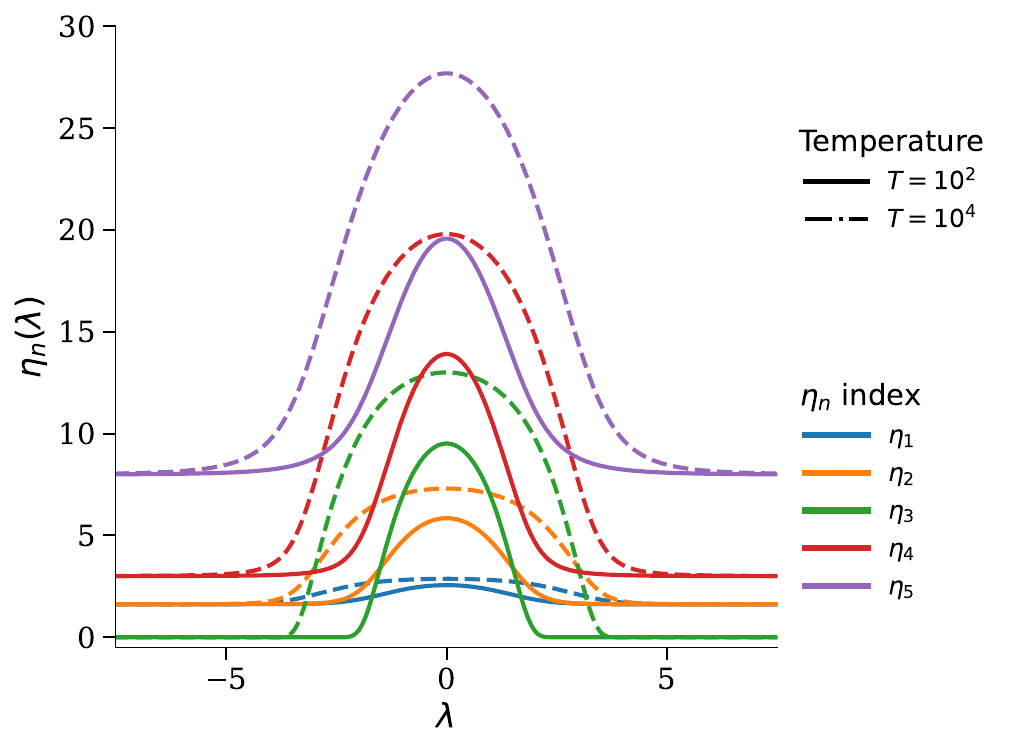}
  \caption{
  Representative $\eta_p(\lambda)$ functions for small string indices $p$ in the two-channel ($n=2$) 
  and three-channel ($n=3$) Gross--Neveu models. 
  In both cases, $\eta_p(\lambda)$ decreases monotonically with increasing $p$, and the sequence terminates at the channel number, 
  such that $\eta_{p=n}(\lambda)\to 0$. 
  This behavior reflects the exact truncation of the $\eta$-string hierarchy at $p=n$, characteristic of the $n$-channel impurity problem.}
  \label{fig:eta_n_2ch_3ch}
\end{figure*}

\subsection{The overscreened Kondo phase}
For $ d$ real or purely imaginary, $d = i\delta$ with $\delta \in (0, \frac{1}{2})$  corresponding to $J\gg g$.  As indicated by the perturbative one-loop beta function
\begin{equation}
    \frac{\pi}{2}  \beta_g(J) = -J^2 + gJ,
\end{equation}
 the beta function of the boundary coupling $J$ is negative, signaling that the weak-coupling fixed point is unstable and the Hamiltonian flows away from it under renormalization. The perturbative RG analysis alone does not determine the ultimate fixed point; however, the exact Bethe Ansatz solution shows that the flow terminates at an intermediate, non-Fermi-liquid fixed point. Around this fixed point, the impurity is overscreened by a multiparticle cloud of gapped bulk quasiparticles, with the screening characterized by a Kondo scale $T_K=f(d)m$ with $f(d)$ being a function of the RG-invariant parameter $d$ (See Appendix~\ref{RGderivation-Nch}). Deep in the Kondo phase, for $d\gg 1$ we have $f(d)\to e^{\pi d}$ with the scale taking  its canonical form $T_K= De^{-\pi/J}$. Over most of the Kondo phase $T_K \gg 2m$, with the Kondo scale dominating over the bulk superconducting gap.

 The Kondo scale determines the thermodynamics of the impurity in the Kondo phase. For $T<T_K$ the impurity is overscreened, while for $T>T_K$ it is asymptotically free. This overscreening mechanism parallels that of the conventional multichannel Kondo effect with gapless conduction electrons, but here it arises in a superconducting environment with a finite bulk gap where the overscreening cloud is formed by gapped bulk quasiparticles.

The crossover between the asymptotically free impurity at high temperatures and the overscreened impurity at low temperatures  requires the solution of the thermodynamic Bethe Ansatz  (TBA) equations. These equations take the following form ~\cite{andrei1984solution,PhysRevB.58.7619}
    \begin{equation}\label{TBAg}
    \ln \eta_p(\lambda) = 
   -\frac{m}{T}\cosh(\pi \lambda)\delta_{p,n}+\sum_{\upsilon=\pm 1}G\ln\left[ 1+\eta_{p+\upsilon } \right]
\end{equation}
where
\begin{equation}
    Gf(\lambda) = \int \mathrm{d} \mu  \frac{1}{2\cosh\pi (\lambda-\mu)}f(\mu).
\end{equation}

To close these equations, we need to supply boundary conditions at $p\to \infty$, which gives a new relation
    \begin{equation}
    \lim_{p\rightarrow \infty} \left\{[p+1]\ln(1+\eta_p(\mu))-[p]\ln(1+\eta_{p+1}(\mu))\right\}=-\frac{H}{T}
    \label{tbainftybc}
\end{equation}
where $[p]$ is an integral kernel $\pi^{-1}(p\pi/2)[(p\pi/2)^2 + \mu^2]^{-1}
$, and $H$ is the applied global magnetic field. Moreover, we choose the boundary conditions at $0$ as $\eta_0(\lambda)=0$.

These thermodynamic Bethe Ansatz equations are identical in all four phases of the model, but the specific form of the functions $\eta_p(T,\lambda)$ depends on the number of channels $n$. 
For $n=2$ and $n=3$, several representative $\eta_p(T,\lambda)$ with $p=1,\dots,5$ are shown in Fig.~\ref{fig:eta_n_2ch_3ch}.
Although the equations themselves do not change across phases, the way in which the impurity free energy is composed from the $\eta_p$ differs between phases, leading to distinct thermodynamic behavior.

The impurity part of the free energy in the overscreened Kondo phase (derived following the method discussed in Ref.~\cite{kattel2025thermodynamics}) is
\begin{equation}
    \begin{aligned}
        \mathcal{F}_{\mathrm{imp}}&=\mathcal{F}_{\mathrm{imp}}^0-\frac{T}{2}\int \mathrm{d} \lambda \sum_{\upsilon=\pm}\frac{\ln(1+\eta_1(\lambda))}{2\cosh\pi (\lambda+\upsilon d)}.
    \end{aligned}
    \label{freeenegg1}
\end{equation}

The TBA equations (Eq.\ \eqref{TBAg}, with the boundary conditions $\eta_{0}(\lambda)=0$ and Eq.\ \eqref{tbainftybc}) involve the mass scale $m$ explicitly, so no closed‐form solution exists at arbitrary temperature $T$; one must therefore solve them numerically (cf.\ Ref.\ \cite{PhysRevLett.49.497}). However in the two asymptotic limits ($T\to 0$ and $T\to\infty$), the factor $(m/T)\cosh(\pi\lambda)$ either vanishes or diverges, making all $\eta_{p}(\lambda)$ independent of $\lambda$ and greatly simplifying the recursion relation, allowing us to obtain analytic results.

\begin{figure*}
    \centering
    \includegraphics[width=\linewidth]{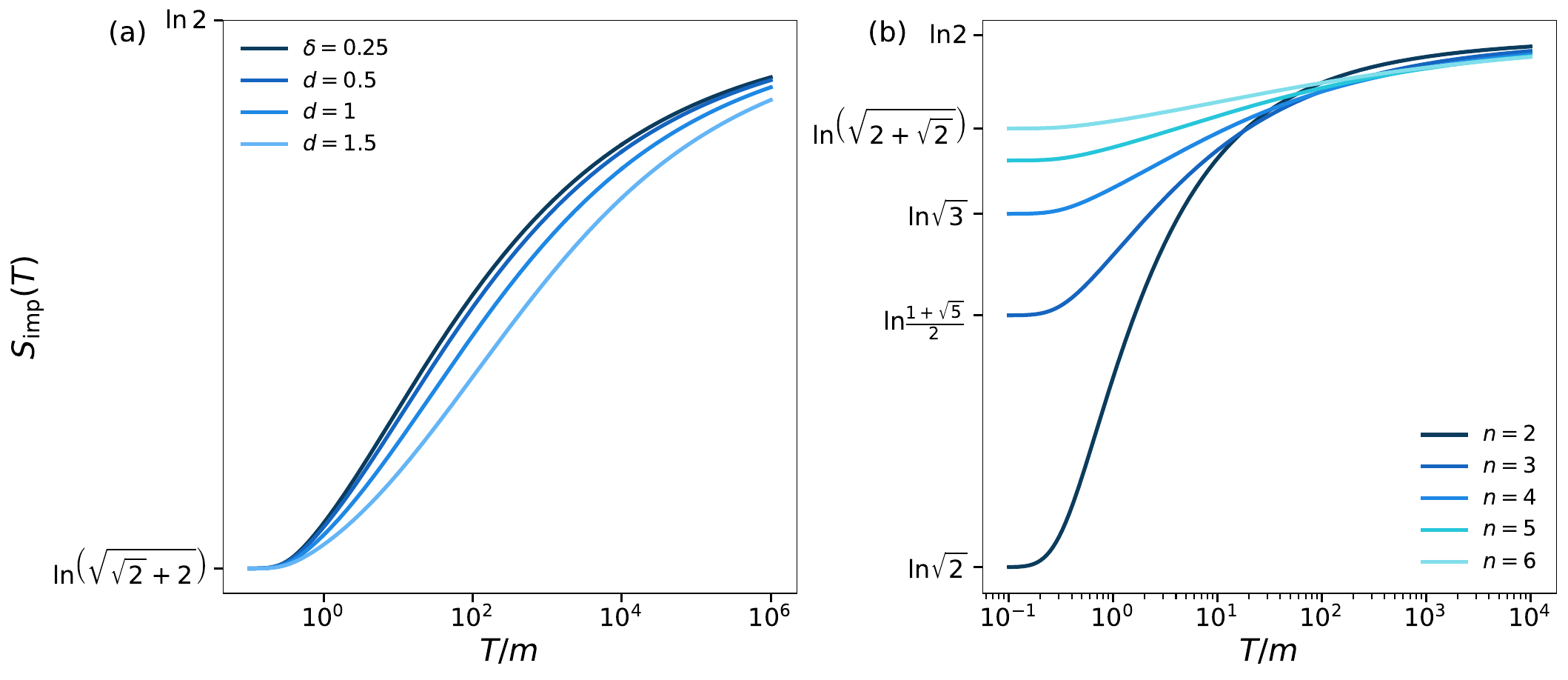}
    \caption{Impurity entropy $S_{\mathrm{imp}}$ in the overscreened Kondo phase as a function of the universal scaling variable $T/m$.
(a) Representative behavior of the impurity entropy for the six-channel superconducting bulk. Colors denote the RG-invariant parameter $d$, demonstrating the universal scaling relation $S_{\mathrm{imp}} = S_{\mathrm{imp}}(d, T/m)$, which takes the value $S_{\mathrm{imp}}(T\to 0)=\ln(\sqrt{2+\sqrt{2}})$ as $T$ approaches zero and asymptotically reaches to $S_{\mathrm{imp}}(T\to \infty)=\ln 2$ at infinite temperature.
(b) For the representative impurity parameter $d=0.25$, $S_{\mathrm{imp}}(T)$ is shown for different numbers of superconducting channels $n$, exhibiting the crossover from $S_{\mathrm{imp}}(0)=\ln \bigl[2\cos \bigl(\frac{\pi}{n+2}\bigr)\bigr]$ to $S_{\mathrm{imp}}(\infty)=\ln 2$.}
    \label{Fig:TBAK}
\end{figure*}

In the absence of an applied magnetic field, consider first  the high temperature regime:  when $T\to\infty$,  $m/T\to 0$ and the driving term vanishes, such that the recursion relation Eq.\eqref{TBAg} becomes completely homogeneous, which leads to a simple solution of the form
\begin{equation}
  {\eta_{p}}_{(T\to\infty)} = (p+1)^{2} - 1,\qquad p = 1,2,\dots.
  \label{inftemplimit}
\end{equation}
Since each ${\eta_{p}}_{(T\to \infty)}$ is constant in $\lambda$, substituting into Eq.\ \eqref{freeenegg1} one obtains that the impurity entropy at infinite temperature approaches 
\begin{equation}
  S_{\rm imp}(T\to\infty) =\frac{1}{2} \ln\left(1+{\eta_{1}}_{(T\to\infty)}\right) = \ln2,
\end{equation}
which shows that the impurity is unscreened, indicating the theory is asymptotically free in the UV, as we already observed from the RG flows

Conversely, in the zero‐temperature limit $T\to0$, the factor $(m/T)\cosh(\pi\lambda)$ diverges and forces $\eta_{n}(\lambda)=0$. For each level $p<n$, the recursion relation Eq.\eqref{TBAg} then reduces to
$ \ln\eta_{p} 
  = 
  \frac12\ln\left(1+\eta_{p-1}\right) 
  + 
  \frac12\ln\left(1+\eta_{p+1}\right),$
whose unique solution is
\begin{equation}
  {\eta_{p}}_{(T\to 0)} 
  =\begin{cases} 
  \frac{\sin^{2}\left(\frac{(p+1)\pi}{n+2}\right)}
       {\sin^{2}\left(\frac{\pi}{n+2}\right)} - 1,
  \qquad p=1,\dots,n-1.\\
(p+1-n)^2-1, ~\quad p=n,\dots
  \end{cases}
  \label{zerotemplimit}
\end{equation}
Once again from Eq.\eqref{freeenegg1}, we find that for constant $\eta_1(\lambda)$, we obtain $S_{\rm imp}=\frac{1}{2}\ln\left(1+{\eta_1}_{(T\to 0)}\right)$ which leads to
\begin{equation}
  S_{\rm imp}(T \to 0) = \ln\left(2\cos\left(\frac{\pi}{n+2}\right)\right).
\end{equation}

This shows that the system flows to the overscreened $n$-channel Kondo fixed point, exhibiting non-Fermi-liquid behavior with residual impurity entropy
\begin{equation}
S_{\rm imp}(T{=}0) = \ln \mathfrak{g}, \quad \mathfrak{g} = 2\cos\left(\frac{\pi}{n+2}\right), \quad n \ge 1,
\end{equation}
corresponding to a noninteger ground state degeneracy. This matches the logarithmic quantum dimension of the spin-$j$ primary in the $SU(2)_n$ WZW conformal field theory. The modular $S$-matrix element
\begin{equation}
\mathsf{S}_{[0,j]} = \frac{2}{n+2} \sin\left( \frac{(j+1)\pi}{n+2} \right)
\end{equation}
yields the impurity entropy as $S_{\rm imp} = \ln \left( \frac{\mathsf{S}_{[0,j]}}{\mathsf{S}_{[0,0]}} \right)$. The impurity provides a boundary condition labeled by a representation of ${SU}(2)_n$, and the IR theory supports chiral parafermionic operators with non-Abelian fusion rules. 

Remarkably, these topological features—and the associated breakdown of Fermi-liquid behavior—persist even when bulk interactions are present ($g \neq 0$) and a gap opens in the bulk. The critical exponents remain
\begin{equation}
\alpha = -\frac{4}{2+n}, \quad \delta = \frac{n}{2}, \quad \forall   n > 2,
\end{equation}
demonstrating the robustness of fractionalized impurity degrees of freedom governed by the same CFT fusion algebra despite bulk perturbations~\cite{kattelGNTBA}.

Having identified the two boundary RG fixed points—the UV limit where the impurity decouples from the bulk and the IR limit where it is overscreened by gapped bulk quasiparticles, yielding non-Fermi-liquid behavior with a finite residual impurity entropy—we now turn to the crossover between these regimes. To quantify this interpolation, we compute the temperature-dependent impurity entropy by numerically solving the (TBA) equations. Representative results for $ n = 6 $ are shown in Fig.~\ref{Fig:TBAK}(a). 
For $ d = 0.25 $, Fig.~\ref{Fig:TBAK}(b) displays the impurity entropy for 
$ n $-channel systems with $ n \in [2,6] $.  From the form of the TBA equations Eq.\eqref{TBAg} and hence from Fig.\ref{Fig:TBAK}, it is evident that the impurity entropy $S_{\rm imp}(d,\frac{m}{T})$ is a universal function of of the scaling variable $m/T$ (and the RG-invariant parameter d). Note that the impurity entropy is a monotonic function of temperature, which smoothly grows from $S_{\rm imp}=\ln \left(2 \cos\left(\frac{\pi}{n+2}\right)\right)$ at $T=0$ to $\ln 2 $ at $T=\infty$ for any $n$ and parameter $d$ either real or purely imaginary $d=i\delta$ with $0\leq \delta<\frac{1}{2}$.

After our discussion of the Kondo phase, we proceed to discuss the remaining three phases and briefly outline their ground‐state properties. In the Kondo phase, as mentioned earlier, all the solutions of the Bethe Ansatz equations take the form of the string solution explicitly written in Eq.\eqref{string-soln}. However, as $\delta$ increases beyond $\frac{1}{2}$, the Bethe Ansatz equations admit a new type of purely imaginary solution of the form $\lambda_\delta= i\left(\delta-\frac{1}{2}\right)$ often called boundary string solution in the literature~\cite{kapustin1996surface,zhang2014exact,skorik1995boundary,pasnoori2022rise,kattel2024overscreened}. Apart from this fundamental boundary string solution, there are also higher order boundary string solutions of the form $\lambda_{\delta,\ell}^{(p)}=\lambda_\delta+i\ell$ with $\ell=1,\dots,p$ and $p=\lfloor\delta+\frac{1}{2}\rfloor$ (where $\lfloor\cdot\rfloor$ is the floor function) depending on the value of the parameter $\delta$~\cite{zhakenov2025thermodynamics,kattel2025thermodynamics}.   As the RG-invariant parameter $\delta$ varies in this system, the excitation described by the boundary-string solution fundamentally changes, which gives rise to distinct ground state behaviors and reveals additional phases. This very solution also complicates the thermodynamic Bethe Ansatz analysis because the Hilbert space splits into different towers~\cite{zhakenov2025thermodynamics,kattel2025thermodynamics}. Thus, for all remaining three phases, we explain the structure of the eigenstate towers, and then we perform the thermodynamic Bethe Ansatz analysis by summing over the tower structure by generalizing the tower summed method recently developed in Refs.~\cite{zhakenov2025thermodynamics,kattel2025thermodynamics}.

\subsection{The Zero-mode phase}

When the RG-invariant parameter $\delta$ is in the range  $\frac{1}{2} < \delta < \frac{n}{2}$, the impurity lies in the zero-mode phase. In this phase, $g\lesssim J$, the strength of the boundary Kondo coupling and the bulk superconducting coupling starts to become comparable. As a result, while the impurity remains screened by the Kondo effect, as in the overscreened phase, the bulk superconducting order stabilizes a subgap excitation. However, unlike the conventional finite mass YSR bound state, this excitation has vanishing energy in the thermodynamic limit. The ground state is still composed of $n$-strings, and the impurity remains overscreened by a Kondo multiparticle cloud, similar to the overscreened Kondo phase described above. However, the uniqueness of this phase is characterized by the existence of a boundary string solution $\lambda_\delta=i\left(\delta-\frac{1}{2}\right)$, which gives rise to a sub-gap zero-energy excitation. This sub-gap excitation is constructed by adding an $(n-1)$-string solution and the boundary string solution on top of the $n$-string sea. This excitation has exactly vanishing energy in the thermodynamic limit, distinguishing it from other phases.

Unlike in the Kondo phase, where the absence of a boundary string yields a single, continuous tower of eigenstates constructed upon the overscreened ground state, the existence of a purely imaginary boundary string $\lambda_d$ and higher-order boundary roots in the zero-mode phase divides the spectrum into two or three separate towers. The number of towers depends on the value of the impurity parameter $\delta$. For Zero-mode Phase I ($\frac{1}{2}<\delta<1$), there are two towers corresponding respectively to configurations where the boundary mode is \emph{unoccupied} or \emph{occupied}, whereas for Zero-mode Phase II ($1<\delta<\frac{n}{2}$), higher order boundary strings of the form $\lambda_{\delta,\ell}^{(p)}=\lambda_\delta+i\ell$ with $\ell=1,\dots,p$ and $p=\lfloor\delta+\frac{1}{2}\rfloor$ (where $\lfloor\cdot\rfloor$ is the floor function) are present such that there are three towers of excitations.

Let us first focus on the Zero-mode Phase I ($\frac{1}{2}<\delta<1$), where the first tower is spanned by the usual bulk string solutions given in Eq.\eqref{string-soln} that contains a total of three-quarters of the states in the total Hilbert space, forming the tower $\mathcal{T}_1$. The second tower, denoted by $\mathcal{T}_2$, comprises the boundary string solution $\lambda_\delta$ together with the associated bulk string configuration, thereby accounting for the remaining quarter of the total states. Together, the two towers $\mathcal{T}_1$ and $\mathcal{T}_2$ span the full Hilbert space in this phase. Then, the impurity contribution to the free energy is expressed as the combined tower sum of the two sectors, given by
\begin{equation}
   e^{-\beta F_{\mathrm{imp}}} = e^{-\beta F_{\rm imp}^{(\mathcal{T}_1)}} + e^{-\beta F_{\rm imp}^{(\mathcal{T}_2)}},
   \label{tower-sum-exp}
\end{equation}
where, 
\begin{align}
    F_{\rm imp}^{(\mathcal{T}_1)}&=F_{\rm imp}^{(\mathcal{T}_2)}-\int \rm{d} \lambda \sum_{\upsilon=\pm}\frac{\frac{T}{4}\ln \left[1+\eta_2(\lambda)\right]}{\cosh(\pi(\lambda+i\upsilon (\delta-\frac12)))}\nonumber\\
		F_{\rm imp}^{(\mathcal{T}_2)}&=\frac{T}{4}\int \rm{d} \lambda \sum_{\upsilon=\pm}\frac{\ln \left[1+\eta_1(\lambda)\right]}{\cosh(\pi(\lambda+i\upsilon \delta))}
		\label{freeengsT}
	\end{align}

Since the dressed energy of the boundary string $\lambda_\delta$ vanishes, none of the towers are lifted in energy.  Using Eq.~\eqref{inftemplimit}, one readily finds that the infinite-temperature impurity entropy contribution due to tower $\mathcal{T}_1$ is 
$S_{\mathrm{imp}}^{(\mathcal{T}_1)}(T \to \infty) = \ln \frac{3}{2}$, 
while the corresponding contribution due to tower $\mathcal{T}_2$ is 
$S_{\mathrm{imp}}^{(\mathcal{T}_2)}(T \to \infty) = \ln \frac{1}{2}$. 
Consequently, the total impurity entropy at infinite temperature is 
$S_{\mathrm{imp}}(T \to \infty) = \ln 2$, as expected. Likewise, using Eq.\eqref{zerotemplimit}, one immediately obtains the zero-temperature contribution to the impurity entropy due to each tower for $n\ge 2$ as
\begin{equation}
    S_{\mathrm{imp}}^{(\mathcal{T}_1)}(T \to 0)
=
\begin{cases}
- \ln \left( 2 \cos \left( \dfrac{\pi}{n+2} \right) \right), & n = 2, \\[1.2ex]
\ln \left( \frac{4 \cos^{2} \left( \dfrac{\pi}{n+2} \right) - 1}{2 \cos \left( \dfrac{\pi}{n+2} \right)} \right), & n \ge 3.
\end{cases}
\label{T1contZM}
\end{equation}
and
\begin{equation}
    S_{\mathrm{imp}}^{(\mathcal{T}_2)}(T \to 0)
= - \ln \left( 2 \cos \left( \frac{\pi}{n+2} \right) \right).
\label{T2contZM}
\end{equation}
Thus, it follows from Eq.\eqref{tower-sum-exp}, that the zero temperature entropy of the impurity is
\begin{equation}
    S_{\rm imp}(T \to 0)=\ln\left(2\cos\left(\frac{\pi}{n+2}\right)\right),
\end{equation}
just as in the Kondo phase. We shall return to a detailed discussion of this result later. For now, we emphasize that it is highly nontrivial: in contrast to the Kondo phase, where the excited eigenstates are organized within a single tower, the present regime features two distinct towers of excitations. The free energy receives separate contributions from these two sectors, each taking a different analytic form. Remarkably, when their contributions are combined with the appropriate statistical weights, the total free energy reproduces the characteristic result of the multichannel Kondo effect---namely, the fractional residual impurity entropy at zero temperature.

 \begin{figure*}
    \centering
    \includegraphics[width=\linewidth]{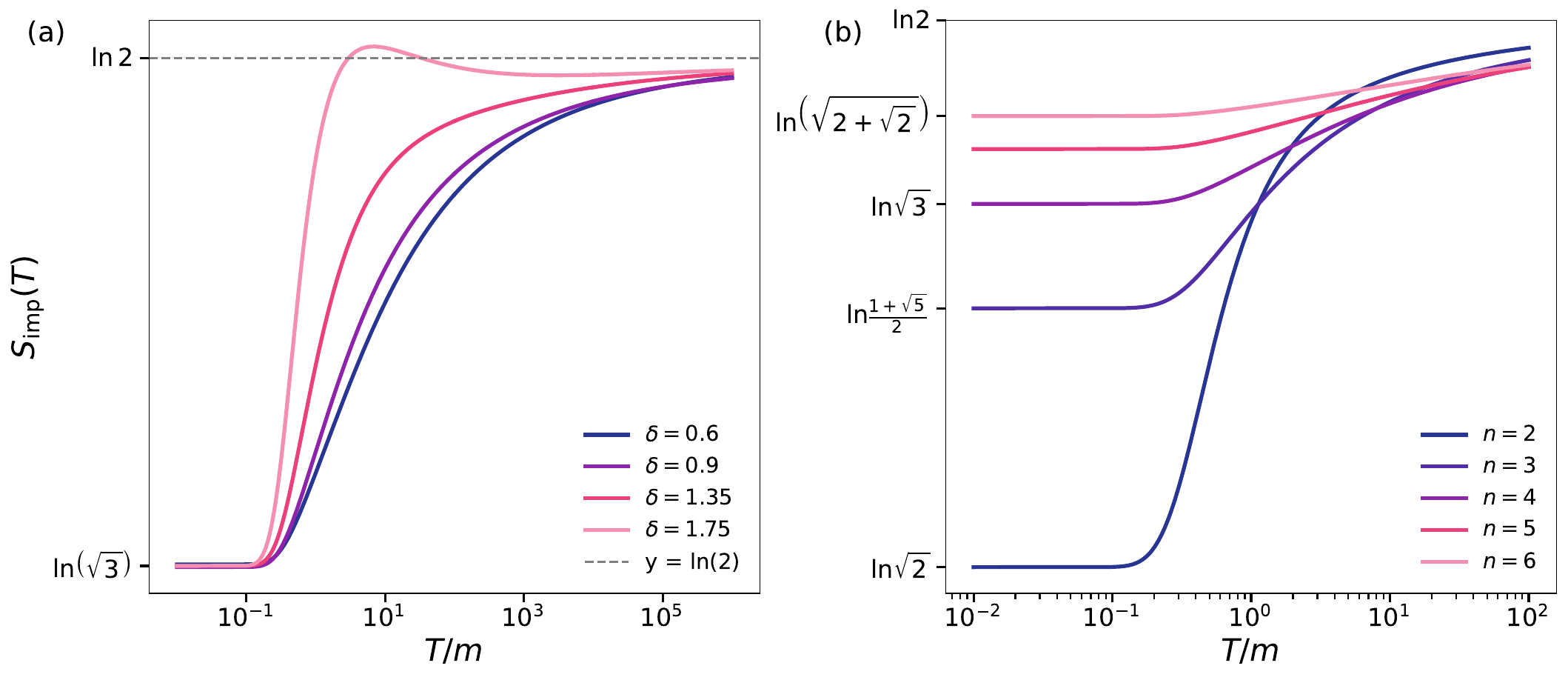}
    \caption{Impurity entropy $S_{\mathrm{imp}}$ in the Zero-mode phase as a function of the universal scaling variable $T/m$.
(a) Representative behavior of the impurity entropy for the four-channel superconducting bulk. Colors denote the RG-invariant parameter $d$, demonstrating the universal scaling relation $S_{\mathrm{imp}} = S_{\mathrm{imp}}(d, T/m)$, which takes the value $S_{\mathrm{imp}}(T\to 0)=\ln\sqrt{3}$ as $T$ approaches zero and asymptotically reaches to $S_{\mathrm{imp}}(T\to \infty)=\ln 2$ at infinite temperature. When $\delta=0.6$ and $\delta=0.9$, the model is in Zero-mode Phase I where there are two towers and hence the impurity is computed from the sum of two towers given by the free energies in Eq.\eqref{T1contZM} and Eq.\eqref{T2contZM}, whereas for $\delta=1.35$ and $\delta=1.75$, the model is in Zero-mode phase II where there are three towers and the free energy contributions from the three towers given in Eq.\eqref{eq:Fimp_all} are used to compute the impurity entropy. 
(b) For the representative impurity parameter $\delta=0.75$, $S_{\mathrm{imp}}(T)$ is shown for different numbers of superconducting channels $n$, exhibiting the crossover from $S_{\mathrm{imp}}(0)=\ln \bigl[2\cos \bigl(\frac{\pi}{n+2}\bigr)\bigr]$ to $S_{\mathrm{imp}}(\infty)=\ln 2$. Here, since $\delta<1$, the impurity entropy comes from the combined contribution of two towers.}
    \label{Fig:TBAB}
\end{figure*}

We shall now discuss the structure of the Hilbert space and the associated impurity thermodynamics for Zero-mode Phase II ($1<\delta<\frac{n}{2}$). In this case, as mentioned earlier, there are three distinct towers of excitations when $\delta>1$. As before, $\mathcal{T}_1$ contains only the usual bulk string solutions given by Eq.\eqref{string-soln}, whereas the tower $\mathcal{T}_2$ contains the fundamental boundary string solution $\lambda_\delta$ and the allowed string configurations. Finally, the third tower $\mathcal{T}_3$ contains both the fundamental string solutions $\lambda_\delta$ and  higher order strings solution $\lambda_\delta+i\ell$ where $\ell=1,\dots,p$ and $p=\lfloor\delta+\frac{1}{2}\rfloor$. The dressed energies of all the boundary string solutions, the fundamental and the higher order ones, vanish in the thermodynamic limit in this regime. Appropriately summing over the eigenstates in each tower following the methods developed in Ref~\cite{kattel2025thermodynamics}, when $\delta>1$, the free energy contribution due to each tower is given by

\begin{subequations}\label{eq:Fimp_all}
\begin{align}
F_{\mathrm{imp}}^{(\mathcal{T}_1)} &= -\frac{T}{4} \sum_{\upsilon = \pm} \int d\lambda 
\left[
\frac{\ln \left(1 + \eta_{\lfloor 2\delta \rfloor +1}(\lambda)\right)}
     {\cosh \left(\pi\left(\lambda + \frac{i\upsilon}{2}(2\delta - \lfloor 2\delta \rfloor)\right)\right)}
\right. \nonumber\\[3pt]
&\left.
- \frac{\ln \left(1 + \eta_{\lfloor 2\delta \rfloor}(\lambda)\right)}
     {\cosh \left(\pi\left(\lambda + \frac{i\upsilon}{2}(\lfloor 2\delta \rfloor+1 - 2\delta)\right)\right)}
\right], \\[6pt]
F_{\mathrm{imp}}^{(\mathcal{T}_2)} &= \frac{T}{4} \sum_{\upsilon = \pm} \int d\lambda 
\left[
\frac{\ln \left(1 + \eta_{\lfloor 2\delta \rfloor - 1}(\lambda)\right)}
     {\cosh \left(\pi\left(\lambda + \frac{i\upsilon}{2}(2\delta - \lfloor 2\delta \rfloor)\right)\right)}
\right. \nonumber\\[3pt]
&\left.
- \frac{\ln \left(1 + \eta_{\lfloor 2\delta \rfloor - 2}(\lambda)\right)}
     {\cosh \left(\pi\left(\lambda + \frac{i\upsilon}{2}(\lfloor 2\delta \rfloor+1 - 2\delta)\right)\right)}
\right], \\[6pt]
F_{\mathrm{imp}}^{(\mathcal{T}_3)} &= \frac{T}{4} \sum_{\upsilon = \pm} \int d\lambda 
\left[
\frac{\ln \left(1 + \eta_{\lfloor 2\delta \rfloor}(\lambda)\right)}
     {\cosh \left(\pi\left(\lambda + \frac{i\upsilon}{2}(\lfloor 2\delta \rfloor+1 - 2\delta)\right)\right)}
\right. \nonumber\\[3pt]
&\left.
+ \frac{\ln \left(1 + \eta_{\lfloor 2\delta \rfloor - 1}(\lambda)\right)}
     {\cosh \left(\pi\left(\lambda + \frac{i\upsilon}{2}(2\delta - \lfloor 2\delta \rfloor)\right)\right)}
\right].
\end{align}
\end{subequations}

so that using Eq.\eqref{inftemplimit}, it is immediate to compute 
$S_{\mathrm{imp}}^{(\mathcal{T}_1)}(T \to \infty) = \ln\left(\frac{\lfloor 2\delta \rfloor + 2}{\lfloor 2\delta \rfloor + 1}\right), \quad
S_{\mathrm{imp}}^{(\mathcal{T}_2)}(T \to \infty) = \ln\left(\frac{\lfloor 2\delta \rfloor - 1}{\lfloor 2\delta \rfloor}\right), \quad
S_{\mathrm{imp}}^{(\mathcal{T}_3)}(T \to \infty) = -\ln\left(\lfloor 2\delta \rfloor (\lfloor 2\delta \rfloor + 1)\right)$. The total impurity entropy at infinite temperature is then $S_{\rm imp}(T\to\infty)=\ln \sum_i e^{S_{\mathrm{imp}}^{(\mathcal{T}_i)}(T \to \infty) }=\ln 2$ as expected.  Whereas the value as $T\to 0$ can be analytically obtained from Eq.\eqref{zerotemplimit} as
\begin{equation}
\begin{aligned}
S_{\mathrm{imp}}^{(\mathcal{T}_1)}(T \to 0) &= \ln\left(\frac{\sin\left(\frac{(\lfloor 2\delta \rfloor + 2)\pi}{n+2}\right)}{\sin\left(\frac{(\lfloor 2\delta \rfloor + 1)\pi}{n+2}\right)}\right), \\[6pt]
S_{\mathrm{imp}}^{(\mathcal{T}_2)}(T \to 0) &= \ln\left(\frac{\sin\left(\frac{(\lfloor 2\delta \rfloor - 1)\pi}{n+2}\right)}{\sin\left(\frac{\lfloor 2\delta \rfloor\pi}{n+2}\right)}\right), \\[6pt]
S_{\mathrm{imp}}^{(\mathcal{T}_3)}(T \to 0) &= -\ln\left(\frac{\sin\left(\frac{(\lfloor 2\delta \rfloor + 1)\pi}{n+2}\right)
\sin\left(\frac{\lfloor 2\delta \rfloor\pi}{n+2}\right)}{\sin^2\left(\frac{\pi}{n+2}\right)}\right).
\end{aligned}
\end{equation}

such that the total impurity entropy, regardless of the value of the RG-invariant parameter in the range $1<\delta<\frac{n}{2}$, is
\begin{equation}
    S_{\mathrm{imp}}(T \to 0)
= \ln\left( \sum_{i=1}^3 e^{S_{\mathrm{imp}}^{(\mathcal{T}_i)}(T \to 0)} \right)
= \ln\left( 2\cos\frac{\pi}{n+2} \right),
\end{equation}
which coincides exactly with the residual impurity entropy of the overscreened $n$-channel Kondo fixed point in the conventional multichannel Kondo problem.

We pause here to emphasize that this is a highly nontrivial result. 
The emergence of a Kondo–like screening structure in this context is far from obvious. 
It arises from a subtle and unexpected interplay between distinct spectral towers, each obeying its own thermodynamic equation for the impurity entropy. 
Individually, none of these towers reproduces the characteristic behavior of the conventional overscreened $n$-channel Kondo fixed point. 
However, when their contributions are combined, they conspire to produce \emph{exactly} the same universal entropy,
\begin{equation}
    S_{\mathrm{imp}}(T \to 0)
= \ln\left(2\cos\frac{\pi}{n+2}\right),
\end{equation}
as that of the overscreened $n$-channel Kondo problem, remarkably, even though the present model possesses a finite spectral gap in the spin sector. 

Thus, in the Zero-mode Phase I i.e. in the parametric regime $\frac{1}{2}<\delta<1$, the impurity entropy originates from the interplay of two such spectral towers. 
Whereas in the Zero-mode Phase II which lies in the parametric range $1<\delta<\frac{n}{2}$, three towers contribute to the impurity entropy, yet their combined effect again yields precisely the same residual entropy as in the gapless multichannel Kondo fixed point. This near-miraculous cancellation underscores the deep universality of the underlying screening mechanism, which transcends the presence or absence of spin-sector gaplessness when the competition between the boundary Kondo scale subtly dominates over the bulk superconductivity, thereby giving rise to this unique zero-mode phase.

Apart from the limiting cases $T=0$ and $T=\infty$, obtaining the impurity entropy in closed analytical form is generally intractable. We therefore compute it numerically by solving the TBA equations. Figure~\ref{Fig:TBAB}(a) presents representative results for the impurity entropy in the four-channel case, plotted as a function of temperature for various values of the RG-invariant parameter $\delta$. Specifically, we show $\delta = 0.6$ and $\delta = 0.9$, corresponding to the regime with two excitation towers, where the respective free-energy contributions are given by Eqs.~(\ref{T1contZM}) and (\ref{T2contZM}), and $\delta = 1.35$ and $\delta = 1.75$, where three excitation towers are present and the corresponding free-energy contributions follow Eq.~(\ref{eq:Fimp_all}). As seen in Fig.~\ref{Fig:TBAB}(a), for values of $\delta$ in the zero-mode regime close to the YSR phase, the impurity entropy can exhibit a non-monotonic dependence on temperature, in contrast to the Kondo phase where it remains strictly monotonic. Due to the presence of a zero-energy subgap state, thermal activation of this mode, together with the smooth entropy crossover from the overscreened regime at low temperatures to the unscreened regime at high temperatures, can lead to a non-monotonic temperature dependence of the impurity entropy. In particular, within certain parameter ranges of the zero-mode phase, the impurity entropy may even exceed the free-moment value of $\ln 2$. Such an effect, where the impurity entropy surpasses $\ln 2$, has already been observed in the single-channel case in regimes exhibiting boundary excitations.
 In Fig.~\ref{Fig:TBAB}(b), we show results for $\delta = 0.75$, illustrating the evolution of the impurity entropy with the number of superconducting leads in the bulk. Since $\delta = 0.75 < 1$, this regime involves only two excitation towers.

\subsection{The YSR regimes}
As $\delta$ further increases, the bulk superconducting order begins to dominate over the Kondo scale. A quantum phase transition occurs at $\delta = \frac{n}{2}$, marking a fundamental change in the impurity screening within the ground state. For values of $\delta$ in the range $\frac{n}{2} < \delta < \frac{n}{2} + 1$, the system enters the YSR phase where the impurity remains unscreened in the ground state, while the excitation described by the fundamental boundary string $\lambda_\delta$ becomes a mid-gap bound state with energy $E = m\left(1 - \cos(\pi(\frac{n}{2} - \delta))\right)$. Thus, one of the characteristic features of this phase is that while the impurity is unscreened in the ground state, it is screened by this single-particle bound mode in the mid-gap state, which is constructed by adding the boundary string solution and one $(n-1)$-string solution of the Bethe Ansatz equations. In the YSR regime, there exist three distinct towers of eigenstates in total, as mentioned earlier. However, the detailed structure of these towers differs between the parameter ranges $ \frac{n}{2} < \delta < \frac{n+1}{2} $ and $ \frac{n+1}{2} < \delta < \frac{n}{2} + 1 $. For clarity, we refer to the former as the \textit{YSR I} regime and the latter as the \textit{YSR II} regime. In particular, due to the presence of massive boundary strings, certain eigenstate towers are lifted in energy. This lifting differs between the YSR I and YSR II regimes. The impurity entropy contribution from each tower in both regimes can be succinctly expressed as
\begin{figure}
    \centering
    \includegraphics[width=0.5\textwidth]{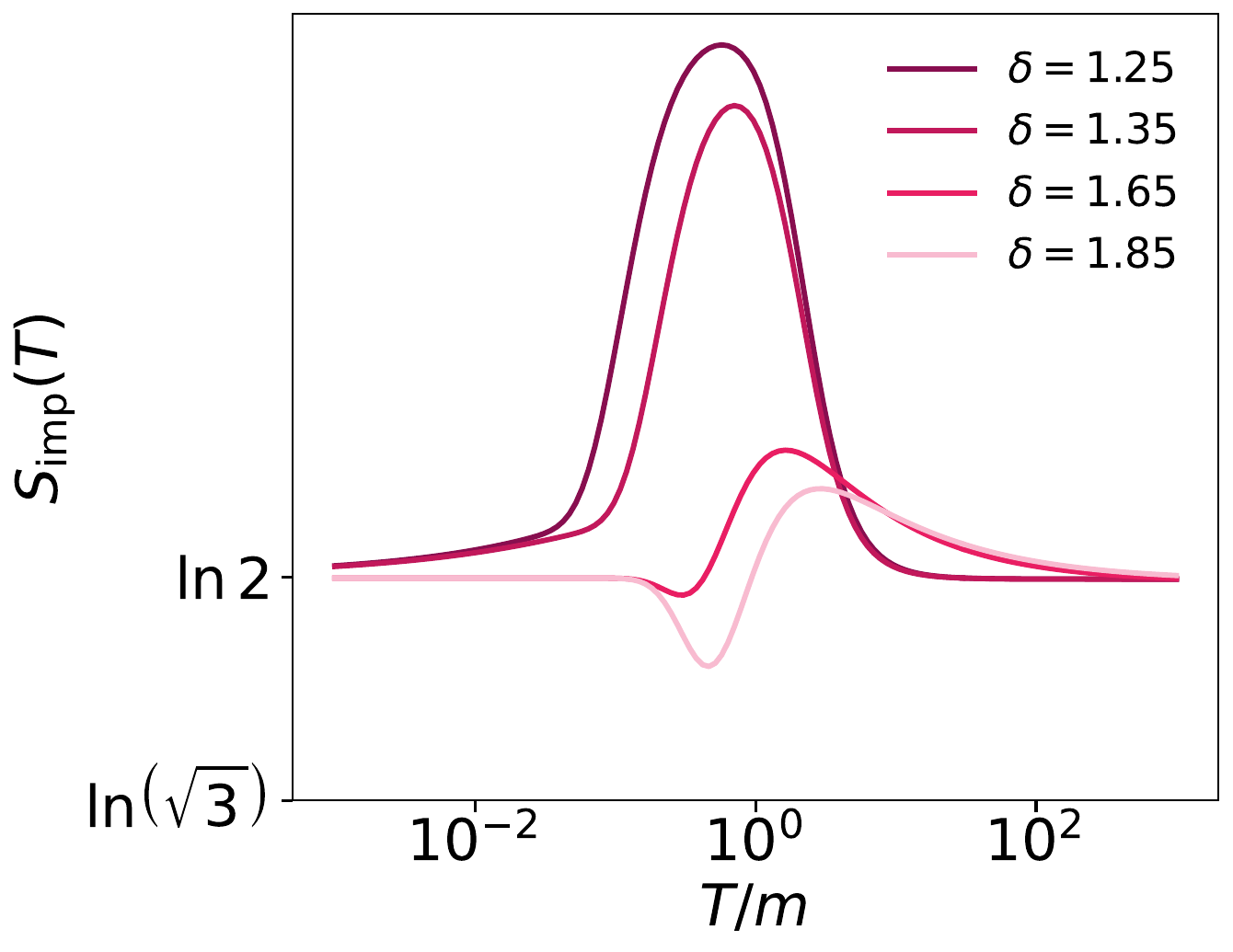}
    \caption{Impurity entropy for the representative two-channel case ($n=2$) plotted as a function of temperature for various values of $\delta$ in the YSR I and YSR II regimes. The impurity entropy approaches $\ln 2$ in both the $T \to 0$ and $T \to \infty$ limits, but exhibits non-monotonic overshoots or undershoots at intermediate temperatures due to the presence of mid-gap YSR states, which become thermally activated only when $T \sim E_\delta$.}
    \label{fig:placeholder}
\end{figure}

\begin{widetext}
    \begin{equation}
\begin{aligned}
F_{\rm imp}^{(\mathcal{T}_1)} &= - \frac{T}{4} \sum_{\upsilon = \pm} \int \mathrm{d}\lambda 
\left[
\frac{\ln \left(1 + \eta_{\lfloor 2\delta \rfloor +1}(\lambda)\right)}
     {\cosh \left(\pi\left(\lambda + \frac{i\upsilon}{2}(2\delta - \lfloor 2\delta \rfloor)\right)\right)}
-
\frac{\ln \left(1 + \eta_{\lfloor 2\delta \rfloor}(\lambda)\right)}
     {\cosh \left(\pi\left(\lambda + \frac{i\upsilon}{2}(\lfloor 2\delta \rfloor+1 - 2\delta)\right)\right)}
\right],\\[0.5em]
F_{\rm imp}^{(\mathcal{T}_2)} &= \frac{T}{4} \sum_{\upsilon = \pm} \int \mathrm{d}\lambda 
\left[
\frac{\ln \left(1 + \eta_{\lfloor 2\delta \rfloor - 1}(\lambda)\right)}
     {\cosh \left(\pi\left(\lambda + \frac{i\upsilon}{2}(2\delta - \lfloor 2\delta \rfloor)\right)\right)}
-
\frac{\ln \left(1 + \eta_{\lfloor 2\delta \rfloor - 2}(\lambda)\right)}
     {\cosh \left(\pi\left(\lambda + \frac{i\upsilon}{2}(\lfloor 2\delta \rfloor+1 - 2\delta)\right)\right)}
\right]
+E_\delta^{\mathcal{T}_2},\\[0.5em]
F_{\rm imp}^{(\mathcal{T}_3)} &= \frac{T}{4} \sum_{\upsilon = \pm} \int \mathrm{d}\lambda 
\left[
\frac{\ln \left(1 + \eta_{\lfloor 2\delta \rfloor}(\lambda)\right)}
     {\cosh \left(\pi\left(\lambda + \frac{i\upsilon}{2}(\lfloor 2\delta \rfloor+1 - 2\delta)\right)\right)}
+
\frac{\ln \left(1 + \eta_{\lfloor 2\delta \rfloor - 1}(\lambda)\right)}
     {\cosh \left(\pi\left(\lambda + \frac{i\upsilon}{2}(2\delta - \lfloor 2\delta \rfloor)\right)\right)}
\right]
+E_\delta^{\mathcal{T}_3}.
\label{Eq:YSRFE}
\end{aligned}
\end{equation}
\end{widetext}

In the YSR I regime, both $ E_\delta^{\mathcal{T}_2} $ and $ E_\delta^{\mathcal{T}_3} $ coincide with the energy of the fundamental boundary string, $ E_\delta $. In contrast, within the YSR II regime, one finds $ E_\delta^{\mathcal{T}_2} = E_\delta $ while $ E_\delta^{\mathcal{T}_3} = 0 $. 
Just as in the portion of the zero-mode phase with $ 1 < \delta < \frac{n}{2} $, one can verify from Eq.~\eqref{inftemplimit} that the total impurity entropy in the infinite-temperature limit, obtained from Eq.~\eqref{Eq:YSRFE}, approaches $ \ln 2 $ in the entire YSR regime $\frac{n}{2}<\delta<\frac{n}{2}+1$, indicating that the impurity becomes asymptotically free, as expected. 

At zero temperature, however, the situation differs qualitatively from the previous regimes. Since not all towers start from the same reference energy, they contribute to the zero temperature entropy differently. In the YSR I regime $\left( \frac{n}{2}<\delta<\frac{n+1}{2}\right)$, both $ \mathcal{T}_2 $ and $ \mathcal{T}_3 $ are lifted by $ E_\delta $, and consequently, only the lowest tower $ \mathcal{T}_1 $ contributes to the zero-temperature impurity entropy. Using Eq.~\eqref{zerotemplimit}, one finds
\begin{equation}
    S_{\mathrm{imp}}(T \to 0) = S_{\mathrm{imp}}^{(\mathcal{T}_1)}(T \to 0) = \ln 2.
\end{equation}
Whereas in the YSR II regime ($\frac{n+1}{2}<\delta<\frac{n}{2}+1$), only $\mathcal{T}_2$ is lifted in energy, and hence the zero temperature contribution comes from the sum of the contributions of the tower $\mathcal{T}_1$ and tower $\mathcal{T}_3$. The contribution from $\mathcal{T}_1$ is $S_{\mathrm{imp}}^{(\mathcal{T}_1)}(T \to 0) = \ln\frac{3}{2}$, and from $\mathcal{T}_3$ it is $S_{\mathrm{imp}}^{(\mathcal{T}_3)}(T \to 0) = -\ln 2$, such that the total impurity entropy is
\begin{equation}
    S_{\mathrm{imp}}(T \to 0) = \ln\left(e^{S_{\mathrm{imp}}^{(\mathcal{T}_1)}} + e^{S_{\mathrm{imp}}^{(\mathcal{T}_3)}}\right)
=\ln 2.
\end{equation}

Thus, in the entire YSR regime, in both the ultraviolet (high-temperature) and infrared (zero-temperature) limits, the impurity remains unscreened, retaining its free-moment entropy $ \ln 2 $. However, at intermediate temperatures $ T \sim E_\delta $, thermal excitation of the mid-gap state with energy $ E_\delta $ renders the higher towers ($ \mathcal{T}_2 $ and $ \mathcal{T}_3 $) thermally accessible. As a result, the impurity entropy is expected to exhibit a non-monotonic dependence on temperature, temporarily exceeding the free-moment value of $ \ln 2 $ before saturating back to it at both asymptotic limits.

\subsection{The unscreened phase}
Finally, for $\delta > \frac{n}{2} + 1$, the system transitions into the local–moment (unscreened) phase. Here, the $\beta$ function $\beta_g(J)$ reverses sign as the superconducting coupling $g$ becomes significantly larger than the Kondo coupling $J$, thereby preventing complete screening of the impurity at any energy scale. All boundary–string solutions possess vanishing energy in the thermodynamic limit, as in the zero–mode phase. Consequently, the impurity–entropy contribution from each tower is again given by the same expression as in the zero–mode phase, namely Eq.~\eqref{eq:Fimp_all}. Since $\delta>\frac{n}{2}+1$, all the $\eta_p$ in Eq.\eqref{eq:Fimp_all} take the value $\eta_{p(T\to 0)}=(p+1-n)^2-1$ according to Eq.\eqref{zerotemplimit} and hence the zero temperature entropy contribution from each tower becomes 
\begin{align}
S_{\mathrm{imp}}^{(\mathcal{T}_1)}(T \to 0)
&= \log\left(\frac{\lfloor 2\delta \rfloor + 2 - n}{\lfloor 2\delta \rfloor + 1 - n}\right),\\
S_{\mathrm{imp}}^{(\mathcal{T}_2)}(T \to 0)
&= \log\left(\frac{\lfloor 2\delta \rfloor - n - 1}{\lfloor 2\delta \rfloor - n}\right),\\
S_{\mathrm{imp}}^{(\mathcal{T}_3)}(T \to 0)
&= -\log\left((\lfloor 2\delta \rfloor + 1 - n)(\lfloor 2\delta \rfloor - n)\right),
\end{align}
such that the zero temperature impurity entropy is $\ln 2$ and, as usual, the infinite temperature total impurity entropy from the sum of contributions from the three towers is also $\ln 2$. 

\begin{figure}[H]
    \centering
    \includegraphics[width=0.5\textwidth]{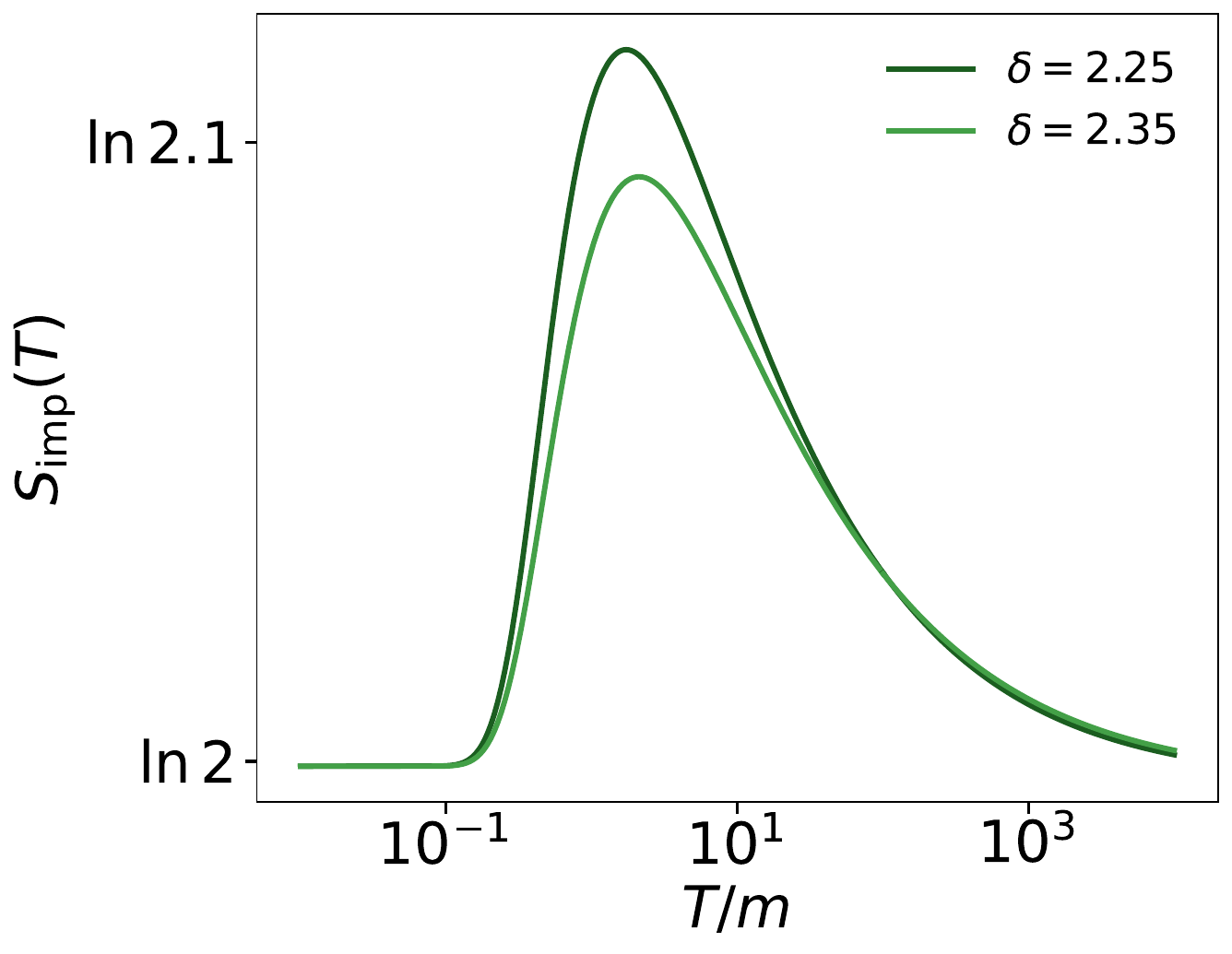}
    \caption{Representative plots of the impurity entropy for $n=2$ and the parameter $\delta>\frac{n}{2}+1$ such that the model is in the unscreened phase. The impurity entropy is $\ln 2$ both in the UV and IR with non-monotonic humps in the intermediate temperature regime, which progressively vanish as $\delta\to\infty$. }
    \label{fig:US2ch}
\end{figure}

As shown in Fig.~\ref{fig:US2ch}, the impurity entropy approaches $\ln 2$ in both the zero-temperature and high-temperature limits. At intermediate temperatures, a pronounced bump appears, where the entropy exceeds the free-moment value $\ln 2$. This feature reflects the competition between the boundary Kondo correlations and the bulk superconducting order. Deep in the unscreened phase, i.e., as $\delta \to \infty$, the superconducting order in the bulk completely dominates over the boundary Kondo effect, the bump progressively disappears, and the entropy remains close to $\ln 2$ across all temperature scales.
\section{Conclusion}
We have constructed and exactly solved a novel integrable model consisting of a single spin-$\frac{1}{2}$ impurity coupled isotropically to $n$ superconducting leads. Our analysis uncovers four distinct boundary phases: the conventional overscreened Kondo phase (which would occupy the entire phase space in the absence of bulk superconducting order) and three additional phases induced by superconductivity, where the transitions are governed by changes in the structure of the boundary towers of excitations.

In the Kondo phase, the impurity is overscreened by a many-body cloud of gapped spinons; remarkably, however, the critical exponents coincide with those of the conventional multichannel Kondo problem featuring gapless spin excitations. In the zero-mode phase, the impurity remains overscreened, but boundary excitations with vanishing energy emerge, yielding a fractional residual entropy identical to that of the overscreened Kondo fixed point. In contrast, the YSR and local-moment (unscreened) phases correspond to incomplete screening: in the former, partial screening occurs at intermediate temperatures due to a single-particle bound state, whereas in the deep unscreened regime, the impurity remains completely free at all scales.

An interesting future direction is to study the channel–anisotropic generalization of this model, where the couplings between the superconducting leads and the impurity differ across channels~\cite{jerez1996solution,zarand2002thermodynamics}. Moreover, it would be interesting to understand the origin of the residual impurity entropy in the gapped Kondo phase from the boundary CFT perspective by starting from the gapless $SU(2)_n$ WZW theory describing the conventional multichannel Kondo fixed point and perturbing it by a relevant bulk operator $\lambda \mathcal{J}_R^a \mathcal{J}_L^a$ that breaks conformal invariance in the IR by opening the spin gap via dimensional transmutation.

\section{Acknowledgments}
We thank Parmesh Pasnoori, Colin Rylands, Yashar Komijani, and Andreas Gleis for valuable and insightful discussions. PK acknowledges the hospitality of the Flatiron Institute, where part of this work was completed.  

\bibliography{ref.bib}

@article{NozieresBlandin1980,
  author  = {Nozières, P. and Blandin, A.},
  title   = {Kondo Effect in Real Metals: Multichannel Kondo Problem},
  journal = {J. Physique},
  volume  = {41},
  pages   = {193--211},
  year    = {1980}
}

@article{AffleckLudwig1993,
  author  = {Affleck, Ian and Ludwig, Andreas W.~W.},
  title   = {Exact Conformal‐Field‐Theory Results on the Multichannel Kondo Effect: Single‐Fermion Green’s Function, Self‐Energy, and Resistivity},
  journal = {Phys. Rev. B},
  volume  = {48},
  pages   = {7297--7321},
  year    = {1993}
}

@article{amaricci2025engineering,
  title={Engineering the Kondo impurity problem with alkaline-earth atom arrays},
  author={Amaricci, Adriano and Richaud, Andrea and Capone, Massimo and Oppong, Nelson Darkwah and Scazza, Francesco},
  journal={arXiv preprint arXiv:2505.14630},
  year={2025},
url={https://arxiv.org/abs/2505.14630}
}

@article{taie2010,
  title={Realization of a SU (2)$\times$ SU (6) system of fermions in a cold atomic gas},
  author={Taie, Shintaro and Takasu, Yosuke and Sugawa, Seiji and Yamazaki, Rekishu and Tsujimoto, Takuya and Murakami, Ryo and Takahashi, Yoshiro},
  journal={Physical review letters},
  volume={105},
  number={19},
  pages={190401},
  year={2010},
  publisher={APS},
url={https://journals.aps.org/prl/abstract/10.1103/PhysRevLett.105.190401}
}

@article{scazza2014,
  title={Observation of two-orbital spin-exchange interactions with ultracold SU (N)-symmetric fermions},
  author={Scazza, Francesco and Hofrichter, Christian and H{\"o}fer, Moritz and De Groot, PC and Bloch, Immanuel and F{\"o}lling, Simon},
  journal={Nature Physics},
  volume={10},
  number={10},
  pages={779--784},
  year={2014},
  publisher={Nature Publishing Group UK London},
url={https://www.nature.com/articles/nphys3061}
}

@article{Pagano2014,
  author    = {Pagano, Guido and Mancini, Marco and Cappellini, Giacomo and Lombardi, Pietro and Schäfer, Florian and Hu, Hui and Liu, Xia-Ji and Catani, Jacopo and Sias, Carlo and Inguscio, Massimo and Fallani, Leonardo},
  title     = {A one-dimensional liquid of fermions with tunable spin},
  journal   = {Nature Physics},
  volume    = {10},
  number    = {3},
  pages     = {198--201},
  year      = {2014},
  month     = {Mar},
  doi       = {10.1038/nphys2878},
  url       = {https://doi.org/10.1038/nphys2878}
}

@article{nascimbene2013realizing,
  title={Realizing one-dimensional topological superfluids with ultracold atomic gases},
  author={Nascimbene, Sylvain},
  journal={Journal of Physics B: Atomic, Molecular and Optical Physics},
  volume={46},
  number={13},
  pages={134005},
  year={2013},
  publisher={IOP Publishing},
url={https://iopscience.iop.org/article/10.1088/0953-4075/46/13/134005}
}

@article{andrei1984solution,
  title={Solution of the multichannel Kondo problem},
  author={Andrei, Natan and Destri, C},
  journal={Physical review letters},
  volume={52},
  number={5},
  pages={364},
  year={1984},
  publisher={APS},
url={https://journals.aps.org/prl/abstract/10.1103/PhysRevLett.52.364}
}

@article{tsvelick1984solution,
  title={Solution of the n-channel Kondo problem (scaling and integrability)},
  author={Tsvelick, AM and Wiegmann, PB},
  journal={Zeitschrift f{\"u}r Physik B Condensed Matter},
  volume={54},
  pages={201--206},
  year={1984},
  publisher={Springer},
url={https://link.springer.com/article/10.1007/BF01319184}
}

@article{rylands2016quantum,
  title={Quantum impurity in a Luttinger liquid: Exact solution of the Kane-Fisher model},
  author={Rylands, Colin and Andrei, Natan},
  journal={Physical Review B},
  volume={94},
  number={11},
  pages={115142},
  year={2016},
  publisher={APS},
url={https://journals.aps.org/prb/abstract/10.1103/PhysRevB.94.115142}
}

@article{kattel2024overscreened,
  title={Kondo overscreening in the presence of superconductivity},
  author={Kattel, Pradip and Zhakenov, Abay and Andrei, Natan},
  journal={arXiv preprint arXiv:2412.01924},
  year={2024},
url={https://arxiv.org/abs/2412.01924v2}
}

@article{pasnoori2022rise,
  title={Rise and fall of Yu-Shiba-Rusinov bound states in charge-conserving s-wave one-dimensional superconductors},
  author={Pasnoori, Parameshwar R and Andrei, Natan and Rylands, Colin and Azaria, Patrick},
  journal={Physical Review B},
  volume={105},
  number={17},
  pages={174517},
  year={2022},
  publisher={APS},
url={https://journals.aps.org/prb/abstract/10.1103/PhysRevB.105.174517}
}

@article{Yu,
author = {YU LUH},
title = {BOUND STATE IN SUPERCONDUCTORS WITH PARAMAGNETIC IMPURITIES},
publisher = {Acta Physica Sinica},
year = {1965},
journal = {Acta Physica Sinica},
volume = {21},
number = {1},
eid = {75},
numpages = {16},
pages = {75},
url = {http://wulixb.iphy.ac.cn/EN/abstract/article_851.shtml},
doi = {10.7498/aps.21.75}
}

@article{Shiba,
    author = {Shiba, Hiroyuki},
    title = "{Classical Spins in Superconductors}",
    journal = {Progress of Theoretical Physics},
    volume = {40},
    number = {3},
    pages = {435-451},
    year = {1968},
    month = {09},
    issn = {0033-068X},
    doi = {10.1143/PTP.40.435},
    url = {https://doi.org/10.1143/PTP.40.435}
}

@ARTICLE{Rusinov,
   author = {{Rusinov}, A.~I.},
    title = "{Superconductivity near a Paramagnetic Impurity}",
  journal = {Soviet Journal of Experimental and Theoretical Physics Letters},
     year = 1969,
    month = jan,
   volume = 9,
    pages = {85},
   adsurl = {https://ui.adsabs.harvard.edu/abs/1969JETPL...9...85R},
url={http://www.jetpletters.ru/ps/1658/article_25295.shtml}
}

@article{andrei1995fermi,
  title={Fermi-and non-fermi-liquid behavior in the anisotropic multichannel kondo model: Bethe ansatz solution},
  author={Andrei, Natan and Jerez, Andr{\'e}s},
  journal={Physical review letters},
  volume={74},
  number={22},
  pages={4507},
  year={1995},
  publisher={APS},
url={https://journals.aps.org/prl/abstract/10.1103/PhysRevLett.74.4507}
}

@article{jerez1998solution,
  title={Solution of the multichannel Coqblin-Schrieffer impurity model and application to multilevel systems},
  author={Jerez, Andr{\'e}s and Andrei, Natan and Zar{\'a}nd, Gergely},
  journal={Physical Review B},
  volume={58},
  number={7},
  pages={3814},
  year={1998},
  publisher={APS},
url={https://journals.aps.org/prb/abstract/10.1103/PhysRevB.58.3814}
}

@article{zinn1998generalized,
  title={The generalized multi-channel Kondo model: Thermodynamics and fusion equations},
  author={Zinn-Justin, P and Andrei, N},
  journal={Nuclear Physics B},
  volume={528},
  number={3},
  pages={648--682},
  year={1998},
  publisher={Elsevier},
url={https://www.sciencedirect.com/science/article/abs/pii/S0550321398003873}
}

@article{PhysRevLett.49.497,
  title = {Thermodynamics of the Kondo Model},
  author = {Rajan, V. T. and Lowenstein, J. H. and Andrei, N.},
  journal = {Phys. Rev. Lett.},
  volume = {49},
  issue = {7},
  pages = {497--500},
  numpages = {0},
  year = {1982},
  month = {Aug},
  publisher = {American Physical Society},
  doi = {10.1103/PhysRevLett.49.497},
  url = {https://link.aps.org/doi/10.1103/PhysRevLett.49.497}
}

@article{PhysRevB.58.7619,
  title = {Chiral liquids in one dimension: A non-Fermi-liquid class of fixed points},
  author = {Andrei, Natan and Douglas, Michael R. and Jerez, Andr\'es},
  journal = {Phys. Rev. B},
  volume = {58},
  issue = {12},
  pages = {7619--7625},
  numpages = {0},
  year = {1998},
  month = {Sep},
  publisher = {American Physical Society},
  doi = {10.1103/PhysRevB.58.7619},
  url = {https://link.aps.org/doi/10.1103/PhysRevB.58.7619}
}

@article{kapustin1996surface,
  title={Surface excitations and surface energy of the antiferromagnetic XXZ chain by the Bethe ansatz approach},
  author={Kapustin, A and Skorik, S},
  journal={Journal of Physics A: Mathematical and General},
  volume={29},
  number={8},
  pages={1629},
  year={1996},
  publisher={IOP Publishing},
url={https://iopscience.iop.org/article/10.1088/0305-4470/29/8/011}
}

@article{zhang2014exact,
  title={Exact solution of the one-dimensional super-symmetric t--J model with unparallel boundary fields},
  author={Zhang, Xin and Cao, Junpeng and Yang, Wen-Li and Shi, Kangjie and Wang, Yupeng},
  journal={Journal of Statistical Mechanics: Theory and Experiment},
  volume={2014},
  number={4},
  pages={P04031},
  year={2014},
  publisher={IOP Publishing},
url={https://iopscience.iop.org/article/10.1088/1742-5468/2014/04/P04031}
}

@article{skorik1995boundary,
  title={Boundary bound states and boundary bootstrap in the sine-Gordon model with Dirichlet boundary conditions},
  author={Skorik, Sergei and Saleur, Hubert},
  journal={Journal of Physics A: Mathematical and General},
  volume={28},
  number={23},
  pages={6605},
  year={1995},
  publisher={IOP Publishing},
url={https://iopscience.iop.org/article/10.1088/0305-4470/28/23/014/pdf}
}

@article{tang2025two,
  title = {Two-channel Kondo behavior in the quantum XX chain with a boundary defect},
  author = {Tang, Yicheng and Kattel, Pradip and Pixley, J. H. and Andrei, Natan},
  journal = {Phys. Rev. B},
  volume = {112},
  issue = {2},
  pages = {L020303},
  numpages = {6},
  year = {2025},
  month = {Jul},
  publisher = {American Physical Society},
  doi = {10.1103/t78z-pyfr},
  url = {https://link.aps.org/doi/10.1103/t78z-pyfr}
}

@article{yazdani1997probing,
  title={Probing the local effects of magnetic impurities on superconductivity},
  author={Yazdani, Ali and Jones, BA and Lutz, CP and Crommie, MF and Eigler, DM},
  journal={Science},
  volume={275},
  number={5307},
  pages={1767--1770},
  year={1997},
  publisher={American Association for the Advancement of Science},
url={https://www.science.org/doi/full/10.1126/science.275.5307.1767}
}

@article{franke2011competition,
  title={Competition of superconducting phenomena and Kondo screening at the nanoscale},
  author={Franke, KJ and Schulze, G and Pascual, JI},
  journal={Science},
  volume={332},
  number={6032},
  pages={940--944},
  year={2011},
  publisher={American Association for the Advancement of Science},
url={https://www.science.org/doi/full/10.1126/science.1202204}
}

@article{hatter2015magnetic,
  title={Magnetic anisotropy in Shiba bound states across a quantum phase transition},
  author={Hatter, Nino and Heinrich, Benjamin W and Ruby, Michael and Pascual, Jose I and Franke, Katharina J},
  journal={Nature communications},
  volume={6},
  number={1},
  pages={8988},
  year={2015},
  publisher={Nature Publishing Group UK London},
url={https://www.nature.com/articles/ncomms9988}
}

@article{lee2017scaling,
  title={Scaling of subgap excitations in a superconductor-semiconductor nanowire quantum dot},
  author={Lee, Eduardo JH and Jiang, Xiaocheng and {\v{Z}}itko, Rok and Aguado, Ram{\'o}n and Lieber, Charles M and De Franceschi, Silvano},
  journal={Physical Review B},
  volume={95},
  number={18},
  pages={180502},
  year={2017},
  publisher={APS},
url={https://journals.aps.org/prb/abstract/10.1103/PhysRevB.95.180502}
}

@article{sugawara1968field,
  title={A field theory of currents},
  author={Sugawara, Hirotaka},
  journal={Physical Review},
  volume={170},
  number={5},
  pages={1659},
  year={1968},
  publisher={APS},
url={https://journals.aps.org/pr/abstract/10.1103/PhysRev.170.1659}
}

@article{kattel2024kondo,
  title={Kondo effect in the isotropic Heisenberg spin chain},
  author={Kattel, Pradip and Pasnoori, Parameshwar R and Pixley, JH and Azaria, Patrick and Andrei, Natan},
  journal={Physical Review B},
  volume={109},
  number={17},
  pages={174416},
  year={2024},
  publisher={APS},
url={https://journals.aps.org/prb/abstract/10.1103/PhysRevB.109.174416}
}

@book{jerez1996solution,
  title={Solution of the anisotropic multichannel Kondo model},
  author={Jerez, Andres},
  year={1996},
  publisher={Rutgers The State University of New Jersey, School of Graduate Studies}
}

@article{zarand2002thermodynamics,
  title={Thermodynamics of the anisotropic two-channel Kondo problem},
  author={Zar{\'a}nd, Gergely and Costi, Theo and Jerez, Andres and Andrei, Natan},
  journal={Physical Review B},
  volume={65},
  number={13},
  pages={134416},
  year={2002},
  publisher={APS},
url={https://journals.aps.org/prb/abstract/10.1103/PhysRevB.65.134416}
}

@article{laflorencie2008kondo,
  title={The Kondo effect in spin chains},
  author={Laflorencie, Nicolas and S{\o}rensen, Erik S and Affleck, Ian},
  journal={Journal of Statistical Mechanics: Theory and Experiment},
  volume={2008},
  number={02},
  pages={P02007},
  year={2008},
  publisher={IOP Publishing},
url={https://iopscience.iop.org/article/10.1088/1742-5468/2008/02/P02007}
}

@article{affleck1999logarithmic,
  title={Logarithmic corrections in quantum impurity problems},
  author={Affleck, Ian and Qin, Shaojin},
  journal={Journal of Physics A: Mathematical and General},
  volume={32},
  number={45},
  pages={7815},
  year={1999},
  publisher={IOP Publishing},
url={https://iopscience.iop.org/article/10.1088/0305-4470/32/45/301}
}

@article{kattelGNTBA,
  title={Thermodynamics of the Kondo impurity at the edge of multiple superconducting leads},
  author={Kattel, Pradip and Zhakenov, Abay and Andrei, Natan},
  journal={(in preparation)},
  year={2025},
}

@article{andrei1980diagonalization,
  title={Diagonalization of the kondo hamiltonian},
  author={Andrei, Natan},
  journal={Physical Review Letters},
  volume={45},
  number={5},
  pages={379},
  year={1980},
  publisher={APS}
}

@article{wei2025kondo,
  title={Kondo impurity in an attractive Fermi-Hubbard bath: Equilibrium and dynamics},
  author={Wei, Zhi-Yuan and Shi, Tao and Cirac, J Ignacio and Demler, Eugene A},
  journal={arXiv preprint arXiv:2501.05562},
  year={2025}
}

@article{kattel2025thermodynamics,
  title={Thermodynamics in a split Hilbert space: Quantum impurity at the edge of a one-dimensional superconductor},
  author={Kattel, Pradip and Zhakenov, Abay and Andrei, Natan},
  journal={arXiv preprint arXiv:2508.19330},
  year={2025}
}

@article{kattel2025competing,
  title={Competing color superconductivity and color kondo effect in quark matter},
  author={Kattel, Pradip and Zhakenov, Abay and Andrei, Natan},
  journal={arXiv preprint arXiv:2507.11617},
  year={2025}
}

@article{moca2025spectral,
  title={Spectral properties of fractionalized shiba states},
  author={Moca, C{\u{a}}t{\u{a}}lin Pa{\c{s}}cu and Hajd{\'u}, Csan{\'a}d and D{\'o}ra, Bal{\'a}zs and Zar{\'a}nd, Gergely},
  journal={Physical Review Letters},
  volume={135},
  number={12},
  pages={126502},
  year={2025},
  publisher={APS}
}

@article{manaparambil2025underscreened,
  title={Underscreened Kondo compensation in a superconductor},
  author={Manaparambil, Anand and Moca, Cǎtǎlin Pa{\c{s}}cu and Zar{\'a}nd, Gergely and Weymann, Ireneusz},
  journal={Physical Review B},
  volume={111},
  number={23},
  pages={235433},
  year={2025},
  publisher={APS}
}

@article{moca2021kondo,
  title={Kondo cloud in a superconductor},
  author={Moca, C{\u{a}}t{\u{a}}lin Pa{\c{s}}cu and Weymann, Ireneusz and Werner, Mikl{\'o}s Antal and Zar{\'a}nd, Gergely},
  journal={Physical Review Letters},
  volume={127},
  number={18},
  pages={186804},
  year={2021},
  publisher={APS}
}

@book{kondo2012physics,
  title={The physics of dilute magnetic alloys},
  author={Kondo, Jun},
  year={2012},
  publisher={Cambridge University Press}
}

@book{hewson1997kondo,
  title={The Kondo problem to heavy fermions},
  author={Hewson, Alexander Cyril},
  number={2},
  year={1997},
  publisher={Cambridge university press}
}

@article{tsvelick1983exact,
  title={Exact results in the theory of magnetic alloys},
  author={Tsvelick, AM and Wiegmann, PB},
  journal={Advances in Physics},
  volume={32},
  number={4},
  pages={453--713},
  year={1983},
  publisher={Taylor \& Francis}
}

@article{furusaki1994kondo,
  title={Kondo effect in a Tomonaga-Luttinger liquid},
  author={Furusaki, Akira and Nagaosa, Naoto},
  journal={Physical review letters},
  volume={72},
  number={6},
  pages={892},
  year={1994},
  publisher={APS}
}

@article{lee1992kondo,
  title={Kondo effect in a Luttinger liquid},
  author={Lee, Dung-Hai and Toner, John},
  journal={Physical review letters},
  volume={69},
  number={23},
  pages={3378},
  year={1992},
  publisher={APS}
}

@article{frojdh1995kondo,
  title={Kondo effect in a Luttinger liquid: Exact results from conformal field theory},
  author={Fr{\"o}jdh, Per and Johannesson, Henrik},
  journal={Physical review letters},
  volume={75},
  number={2},
  pages={300},
  year={1995},
  publisher={APS}
}

@article{furusaki2005kondo,
  title={Kondo Problems in Tomonaga--Luttinger Liquids},
  author={Furusaki, Akira},
  journal={Journal of the Physical Society of Japan},
  volume={74},
  number={1},
  pages={73--79},
  year={2005},
  publisher={The Physical Society of Japan}
}

@article{wang1997exact,
  title={Exact solution of the open Heisenberg chain with two impurities},
  author={Wang, Yupeng},
  journal={Physical Review B},
  volume={56},
  number={21},
  pages={14045},
  year={1997},
  publisher={APS}
}

@article{frahm1997open,
  title={The open spin chain with impurity: an exact solution},
  author={Frahm, Holger and Zvyagin, Andrei A},
  journal={Journal of Physics: Condensed Matter},
  volume={9},
  number={45},
  pages={9939},
  year={1997},
  publisher={IOP Publishing}
}

@article{zhakenov2025thermodynamics,
  title={Thermodynamics in a split Hilbert space: Quantum impurity at the edge of the Heisenberg chain},
  author={Zhakenov, Abay and Kattel, Pradip and Andrei, Natan},
  journal={arXiv preprint arXiv:2508.19334},
  year={2025}
}

@article{andrei1980derivation,
  title={Derivation of the chiral Gross-Neveu spectrum for arbitrary SU (N) symmetry},
  author={Andrei, N and Lowenstein, J Ho},
  journal={Physics Letters B},
  volume={90},
  number={1-2},
  pages={106--110},
  year={1980},
  publisher={Elsevier}
}

@article{ludwig1994exact,
  title={Exact conformal-field-theory results on the multi-channel Kondo effect: Asymptotic three-dimensional space-and time-dependent multi-point and many-particle Green's functions},
  author={Ludwig, Andreas WW and Affleck, Ian},
  journal={Nuclear Physics B},
  volume={428},
  number={3},
  pages={545--611},
  year={1994},
  publisher={Elsevier}
}

@article{parcollet1998overscreened,
  title={Overscreened multichannel SU (N) Kondo model: Large-N solution and conformal field theory},
  author={Parcollet, Olivier and Georges, Antoine and Kotliar, Gabriel and Sengupta, Anirvan},
  journal={Physical Review B},
  volume={58},
  number={7},
  pages={3794},
  year={1998},
  publisher={APS}
}

@article{affleck1995conformal,
  title={Conformal field theory approach to the Kondo effect},
  author={Affleck, Ian},
  journal={arXiv preprint cond-mat/9512099},
  year={1995}
}

@article{pereira2025tunneling,
  title={Tunneling spectroscopy of the spinon-Kondo effect in one-dimensional Mott insulators},
  author={Pereira, Rodrigo G and Marquez, Bruno F and Hallberg, Karen and Bauer, Tim and Egger, Reinhold},
  journal={arXiv preprint arXiv:2508.19084},
  year={2025}
}

@article{komijani2020isolating,
  title={Isolating Kondo anyons for topological quantum computation},
  author={Komijani, Yashar},
  journal={Physical Review B},
  volume={101},
  number={23},
  pages={235131},
  year={2020},
  publisher={APS}
}

@article{nozieres1980kondo,
  title={Kondo effect in real metals},
  author={Nozieres, Ph and Blandin, Annie},
  journal={Journal de Physique},
  volume={41},
  number={3},
  pages={193--211},
  year={1980},
  publisher={Soci{\'e}t{\'e} Fran{\c{c}}aise de Physique}
}

\widetext
\newpage

\appendix

In these appendices, we briefly sketch the derivation of the Bethe Ansatz equations and provide explicit solutions to the Bethe Ansatz equations, thereby supplementing the details omitted from the main text.

\section{Bethe Ansatz equations}\label{sec:BAE}
We briefly sketch the construction of the Bethe Ansatz equation in this section. Using the usual method, we compute the 
bare scattering matrix between a bulk particle labeled $j$ and the impurity labeled $0$
\begin{equation}
    S_{j0} = \frac{I_{j0} - i c P_{j0}}{1 - i c}, \quad \text{where} \quad c = \frac{2 J}{1 - \frac{3 J^2}{4}}.
\end{equation}
Here, $I_{j0}$ denotes the identity operator acting on the combined spin space of the particle $j$ and impurity $0$, while $P_{j0} = \frac{1}{2}(I_{j0} + \vec{\sigma}_j \cdot \vec{\sigma}_0)$ is the spin permutation operator exchanging their spin states. The boundary running coupling $c$ encodes the impurity coupling parameter $J$ and controls the strength of the Kondo interaction.

Likewise, the bare scattering matrices between the two bulk particles $i$ and $j$ of opposite chirality and is given by
\begin{equation}
    S_{ij} = \frac{2 i b    I_{ij} + P_{ij}}{2 i b + 1}, \quad \text{with} \quad b = \frac{4 - g^{2}}{8 g}.
\end{equation}
Here, the running coupling constant $b$ depends on the bulk coupling $g$ and parametrizes the interaction strength in the bulk, which stabilizes the superconducting correlations.

In addition to these, another scattering matrix $W_{ij}$ is required to account for the scattering between particles of the \emph{same} chirality. Unlike $S_{ij}$, the $W$-matrix is not directly fixed by the Hamiltonian but rather by the integrability and consistency conditions of the Bethe Ansatz construction. This matrix is simply given by the spin permutation operator,
\begin{equation}
    W_{ij} = P_{ij}.
\end{equation}
This choice ensures the consistency of wavefunction amplitudes when exchanging two particles of the same chirality.

The consistency and integrability of the model are guaranteed by the following algebraic relations among these $S$-matrices:

\begin{align}
W_{jk} W_{ik} W_{ij} &= W_{ij} W_{ik} W_{jk}, \\
S_{jk} S_{ik} W_{ij} &= W_{ij} S_{ik} S_{jk}, \\
S_{j0} S_{ij} S_{i0} W_{ij} &= W_{ij} S_{i0} S_{ij} S_{j0}.
\end{align}

All these scattering matrices and relations can be embedded into the well-known  rational 6-vertex $R$-matrix given by
\begin{equation}
    R_{ij}(\lambda) = \frac{i \lambda    I_{ij} + P_{ij}}{i \lambda + 1}
\end{equation}
where $\lambda\in \mathbb{C}$ is the spectral parameter. The bare scattering matrices can be obtained as special values of $R(\lambda)$:
\begin{align}
    W_{ij} &= R_{ij}(0) = P_{ij},\\
    S_{ij} &= R_{ij}(2 b),\\
    S_{j0} &= K_{j0}(b) = R_{j0}(b + d) R_{j0}(b - d),
\end{align}
where $K(\lambda)$ is the boundary $K$-matrix constructed as a product of $R$-matrices, and $d = \sqrt{b^{2} - \frac{2b}{c} - 1}$ is an RG-invariant parameter that combines the two running coupling constants $b$ and $c$.

Since all the scattering matrices $S^{ij}, S^{j0}, W^{ij}$ satisfy the Yang-Baxter and reflection equations
which ensures the consistency of multi-particle scattering with boundaries, it is immediate to write down the eigenstates of the $Nn$-particle sector ($N$ particles in each $n$ channel) with energy
\begin{equation}
    E = \sum_{j=1}^{Nn} k_j,
\end{equation}
given by
\begin{equation}
    |\{k_j\}\rangle = \sum_{Q, \vec{a}, \vec{\sigma}} \int \theta(x_Q) A_{\vec{a}}^{\vec{\sigma}}[Q] \prod_{j=1}^{Nn} e^{i \sigma_j k_j x_j} \psi_{a_j \sigma_j}^\dagger(x_j) |0\rangle \otimes |a_0\rangle,
\end{equation}
where the sums run over all spin and chirality configurations $\vec{a} = \{a_1,\ldots,a_N,a_0\}$, $\vec{\sigma} = \{\sigma_1,\ldots,\sigma_N,\sigma_0\}$, and orderings $Q \in S_{Nn}$. The function $\theta(x_Q)$ restricts to the ordering $Q$.

Amplitudes related by exchanging the order of particles or flipping the chirality of the rightmost particle scattering off the impurity are connected by the matrices $W^{ij}$, $S^{ij}$, and $S^{j0}$, respectively.

Imposing the open boundary conditions
\begin{equation}
\psi_+(0) = -\psi_-(0), \quad \psi_+(-L) = -\psi_-(-L),
\end{equation}
leads to the quantization condition
\begin{equation}
    e^{-2 i k_j L} A_{\vec{a}}^{\vec{\sigma}}[Q] = (Z_j)_{\vec{a} \vec{a}'}^{\vec{\sigma} \vec{\sigma}'} A_{\vec{a}'}^{\vec{\sigma}'}[Q],
\end{equation}
where the transfer matrix $Z_j$ transporting the $j$-th particle is
\begin{equation}
Z_j = W^{j,j-1} \cdots W^{j,1} S^{j0} S^{j1} \cdots S^{jN} S^{j0} W^{jN} \cdots W^{j,j+1}.
\end{equation}

Using the Yang-Baxter and reflection relations, the transfer matrices commute:
\begin{equation}
    [Z_j, Z_k] = 0 \quad \forall j,k,
\end{equation}
so they can be simultaneously diagonalized.

The transfer matrix $Z_1$ is related to the Monodromy matrix $\Xi_s^{A}(\lambda)$ as $Z_1 = \tau(b) = \rm{Tr}_{A} \Xi^{A}(b)$, where
\begin{equation}
\Xi^{A}(\lambda) = R^{A1}(\lambda + b) \cdots R^{AN}(\lambda + b) R^{A0}(\lambda + d) R^{A0}(\lambda - d)
\times R^{AN}(\lambda - b) \cdots R^{A1}(\lambda - b).
\label{monodromy}
\end{equation}
Here, $A$ represents an auxiliary space which is taken to be that of a spin-$\frac{1}{2}$ particle, and $\rm{Tr}_{A}$ represents the trace over the auxiliary space. Using the properties of the $R$-matrices one can prove that $[\tau(\lambda), \tau(\mu)] = 0$ and by expanding $\tau(\mu)$ in powers of $\mu$, obtain an infinite set of conserved charges which guarantees integrability. By following the functional Bethe Ansatz approach and $T-Q$ relation, we obtain the Bethe equations where the quasi-momenta satisfy
\begin{equation}
    e^{-2 i k_j L} = \prod_{\upsilon=\pm}\prod_{\alpha=1}^{Mn} \left(\frac{(b + \mu_j) +\upsilon \lambda_\alpha + \frac{i}{2}}{(b + \mu_j) +\upsilon \lambda_\alpha - \frac{i}{2}}\right).
    \label{quasimomrell}
\end{equation}
and the spin rapidities satisfy 
\begin{equation}
   \prod_{\upsilon=\pm}\frac{\lambda_j + \upsilon d + \frac{i}{2}}{\lambda_j +\upsilon d - \frac{i}{2}} \prod_{j=1}^{Nn} \left(\frac{\lambda_\alpha +\upsilon (b + \mu_j) + \frac{i}{2}}{\lambda_\alpha +\upsilon (b + \mu_j) - \frac{i}{2}}\right) =\prod_{\upsilon=\pm} \prod_{\beta \neq \alpha}^M \frac{\lambda_\alpha + \upsilon \lambda_\beta + i}{\lambda_\alpha +\upsilon \lambda_\beta - i}.
   \label{combinedeqn2}
\end{equation}

Here, we introduced the inhomogeneity parameter $\mu_j$ to perform the fusion such that the flavor symmetry effectively fuses the spin fields into higher-spin composite operators. This can be achieved by choosing the inhomogeneity parameter of the form
\begin{equation}
    \mu_{j} = i \left( j - \frac{n+1}{2} \right), \quad j = 1, \ldots, n,
\end{equation}
such that the fused Bethe Ansatz equations become

\begin{align}
    e^{-2ik_j L}=\prod_{\alpha=1}^M \prod_{\upsilon=\pm} \frac{b+\upsilon\lambda_\alpha+i \frac{n}{2}}{b+\upsilon\lambda_\alpha-i \frac{n}{2}}
    \label{chareqn}
\end{align}

where the spin rapidities satisfy  

\begin{align}
&\prod_{\upsilon=\pm}\left(\frac{\lambda_\alpha+\nu d+\frac{i}{2}}{\lambda_\alpha+\nu d-\frac{i}{2}} \right)\left( \frac{\lambda_\alpha+\nu b + i \frac{n}{2}}{\lambda_\alpha+\nu b - i \frac{n}{2}}\right)^N=\prod_{\beta\neq \alpha}\frac{\lambda_\alpha-\lambda_\beta+i }{\lambda_\alpha-\lambda_\beta-i }\frac{\lambda_\alpha+\lambda_\beta+i }{\lambda_\alpha+\lambda_\beta-i }.
    \label{spineqn}
\end{align}

We introduced the inhomogeneity parameters manually and carried out the fusion accordingly. In contrast, Refs.~\cite{andrei1984solution,jerez1998solution,zinn1998generalized} systematically incorporate these inhomogeneities through the addition of a second-derivative term in the Hamiltonian. This routine can likewise be adapted to our model, and we could incorporate the inhomogeneity parameter more microscopically.

\section{Detailed solution to the Bethe Ansatz equations}\label{sec:Det-sol-BAE}

The solutions of the Bethe Ansatz equations Eq.\eqref{chareqn} and Eq.\eqref{spineqn} equations, and hence the structure of the ground state and the excitations about the ground state, depend on the value of the RG-invariant parameter $d$. As will be shown when $d$ is real or takes purely imaginary values $d=i\delta$ where $0<\delta<\frac{1}{2}$, the impurity spin is overscreened by the cloud of bulk massive quasiparticles. We solve these equations in these parametric regimes. 

\subsection{Overscreened Kondo Phase}
The ground state of the model is made up of all n-string solutions, where each string is of the form

\begin{equation}
    \left\{\lambda_{\alpha,j}^{(n)}\right\}_{j=1}^n
=
\left\{\Lambda_\alpha + \frac{i}{2}(n+1 - 2j)
 \middle|  j = 1,\dots,n\right\}, \qquad \Lambda_\alpha\in\mathbb{R}.
 \label{nstring-hyp}
\end{equation}

Plugging in these string solutions to the Bethe Ansatz equations Eq.\eqref{chareqn} and Eq.\eqref{spineqn}, we get the Bethe Ansatz equations describing the ground state configurations as
\begin{equation}
    e^{-2 i k_j L} = \prod_{\alpha=1}^M \prod_{\upsilon = \pm 1}\prod_{\xi=1}^n 
\frac{ \left( b + \upsilon \Lambda_\alpha + i \frac{2\xi - 1}{2} \right)
}{
\left( b + \upsilon \Lambda_\alpha - i \frac{2\xi - 1}{2} \right)
}.
\label{cBAEn}
\end{equation}

and

\begin{align}
    &\prod_{\upsilon = \pm 1} \prod_{j=1}^{n} \left(\frac{\Lambda_{\alpha} + \upsilon b + \frac{i}{2}(2j - 1)}{\Lambda_{\alpha} + \upsilon b - \frac{i}{2}(2j - 1)}\right)^N
    \frac{\Lambda_{\alpha} + i \frac{n}{2}}{\Lambda_{\alpha} - i \frac{n}{2}} 
    \prod_{\upsilon = \pm 1} \frac{\Lambda_{\alpha} + \upsilon d + \frac{i n}{2}}{\Lambda_{\alpha} + \upsilon d -\frac{i n}{2}} \\
    &= - \prod_{\beta=1}^{M} \prod_{\upsilon = \pm 1} \frac{\Lambda_{\alpha} + \upsilon \Lambda_{\beta} + i n}{\Lambda_{\alpha} + \upsilon \Lambda_{\beta} - i n}
    \prod_{\xi=1}^{n - 1} \left(\frac{\Lambda_{\alpha} + \upsilon \Lambda_{\beta} + i \xi}{\Lambda_{\alpha} + \upsilon \Lambda_{\beta} - i \xi}\right)^2.
    \label{sbaen}
\end{align}

Taking logarithms on both sides of the equations, Eq.\eqref{cBAEn} becomes
\begin{align}
    k_j=\frac{\pi n_j}{L}+ \frac{1}{L}\sum_{\alpha=1}^M\sum_{\upsilon=\pm}\sum_{\xi=1}^n \tan^{-1}\left( \frac{2(b+\upsilon \Lambda_\alpha)}{2\xi-1}\right),
\end{align}
such that summing over all momenta, we write the equation for the energy eigenstates as
\begin{equation}
    E=\sum_j k_j=\sum_j\frac{\pi n_j}{L}+ D \sum_{\alpha=1}^M\sum_{\upsilon=\pm}\sum_{\xi=1}^n \tan^{-1}\left( \frac{2(b+\upsilon \Lambda_\alpha)}{2\xi-1}\right).
    \label{Efinal-all-strings}
\end{equation}

Likewise, Eq.\eqref{sbaen} becomes
\begin{align}
	&\sum_{\upsilon=\pm}\left(\sum_{j=1}^n N \tan ^{-1}\left( \frac{2(\Lambda_\alpha+\upsilon b)}{2j-1}\right)+\tan ^{-1}\left( \frac{2(\Lambda_\alpha+\upsilon d)}{n}\right) \right)+\tan ^{-1}\left( \frac{2\Lambda_\alpha}{n}\right)\nonumber\\
	&=\sum_{\beta=1}^M\sum_{\upsilon=\pm}\left(\tan ^{-1}\left( \frac{\Lambda_\alpha +\upsilon \Lambda_\beta}{n}\right) +2\sum_{\xi=1}^{n-1} \tan ^{-1}\left(\frac{(\Lambda_\alpha +\upsilon \Lambda_\beta)}{\xi} \right)\right)+ \pi I_\alpha,
\end{align}
where $I_\alpha\in \mathbb{Z}$ are the spin quantum numbers. In the limit $\Lambda_\alpha \to +\infty$, the quantum numbers $I_\alpha$ approach their maximal value
\begin{equation}
I^\infty = n N + \frac{3}{2} - M (2 n - 1).
\end{equation}
This upper bound corresponds to $n N + 1 - M(2 n - 1)$ available slots for the $M$ quantum numbers $I_{\alpha > 0}$. Denoting by $h$ the number of holes (unoccupied quantum numbers), this counting can be rewritten as
\begin{equation}
n N + 1 - M (2 n - 1) = M + h.
\end{equation}
Solving for the number of $n$-strings $M$ as a function of holes $h$, we get
\begin{equation}
M = \frac{N}{2} + \frac{1 - h}{2 n}.
\end{equation}
Since $h$ is a non-negative integer, the maximal number of $n$-strings is $M = \frac{N}{2}$ (assuming $N$ even for convenience), corresponding to one hole $h=1$. This suggests that for the number of roots to be an integer, there has to be 1 hole present in the ground state. Moreover, in the thermodynamic limit, the spacing between consecutive roots $\Delta \Lambda_\alpha = \Lambda_{\alpha+1} - \Lambda_\alpha$ becomes infinitesimal. Defining the root density
\begin{equation}
\rho_n(\Lambda_\alpha) = \frac{I_{\alpha+1} - I_\alpha}{\Delta \Lambda_\alpha},
\end{equation}
we write the Bethe equations for $\Lambda_{\alpha+1}$ and $\Lambda_\alpha$, subtract them, and expand to leading order in $\Delta \Lambda_\alpha$. This yields the integral equation
\begin{equation}
    2 \rho_n(\Lambda) = f(\Lambda)- \sum_{\upsilon=\pm}\int \mathrm d\Lambda' K_n(\Lambda + \upsilon\Lambda')  \rho_n(\Lambda') + \mathcal{O}\left(\frac{1}{N}\right),
    \label{Krootdens}
\end{equation}
with
\begin{align}
    f(\Lambda) &= \sum_{\upsilon=\pm}\left(\sum_{j=1}^n N a_{\frac{2j-1}{2}}(\Lambda+\upsilon b)+a_{\frac{n}{2}}(\Lambda+\upsilon d)\right)+a_{\frac{n}{2}}(\Lambda)-\delta(\Lambda)\\
    a_\gamma(\Lambda) &= \frac{1}{\pi} \frac{\gamma}{\Lambda^2 + \gamma^2}\\
    K_n(\Lambda)&=a_n(\Lambda)+2\sum_{\xi=1}^{n-1} a_{\xi}(\Lambda
    ),
\end{align}
where $\delta(\Lambda)$ is added to remove the trivial root $\Lambda=0$ which leads to non-normalizable eigenfunction. Now, we add a single hole at position $\theta$ to make the total number of roots an integer as discussed above. The root density expressed in Eq.\eqref{Krootdens} can be efficiently solved in Fourier space to obtain the solution of the form

\begin{align}
    \tilde{\rho}_n(\omega)&=\frac{2N \cos(2b)\sum_{j=1}^n e^{-\frac{2j-1}{2}|\omega|}+2\cos(\omega d)e^{-\frac{n}{2}|\omega|}+e^{-\frac{n}{2}|\omega|}-1-2\cos(\omega\theta)}{2(1+e^{-n|\omega|}+\sum_{\xi}^{n-1}2 e^{-\xi|\omega|} )}\nonumber\\
    &=\frac{e^{\frac{n | \omega | }{2}} \left(2N \cos (2 b) e^{-\frac{1}{2} (n-1) | \omega | } \left(e^{n | \omega | }-1\right)+\left(e^{|
   \omega | }-1\right) \left(-e^{\frac{n | \omega | }{2}} (2 \cos (\theta  \omega )+1)+2 \cos (d \omega )+1\right)\right)}{2\left(e^{|
   \omega | }+1\right) \left(e^{n | \omega | }-1\right)}
   \label{rootdensKgs}
\end{align}

Such that the total number of roots in the ground state is $M=n\int\mathrm{d}\Lambda\rho_n(\Lambda)=n\frac{N}{2}$ such that the total spin in the ground state is 
\begin{equation}
    S^z=\frac{nN+1}{2}-M=\frac{1}{2}.
\end{equation}
The spin-$\frac{1}{2}$ observed in the ground state originates from a propagating hole, not from the impurity. The impurity, which initially carries an entropy of $\ln 2$ at the ultraviolet (UV) fixed point, is completely screened at low energies, causing its entropy to vanish. This will be confirmed by an explicit calculation of the thermodynamic entropy.

The change in the root density is due to the presence of a single hole is
\begin{equation}
    \Delta\rho_\theta(\\\omega)=\frac12\left( -\cos (\theta  \omega ) \left(\coth \left(\frac{n | \omega | }{2}\right)+1\right)\tanh \left(\frac{| \omega | }{2}\right)\right)
\end{equation}
and the energy function in the Fourier space takes the form
\begin{equation}
    E(\omega)=D\left(\frac{2i\pi}{\omega}e^{-ib\omega}\right)\frac{e^{\frac{1}{2} (1-2 n) | \omega | } \left(e^{n |
   \omega | }-1\right)}{e^{| \omega | }-1}.
   \label{energyeqn}
\end{equation}
Thus, using the Parseval–Plancherel  theorem, the energy of a single hole becomes
\begin{equation}
    E_\theta=-i D\int\mathrm{d}\omega\frac{ e^{-i b \omega } \rm{sech}\left(\frac{| \omega
   | }{2}\right) \cos (\theta  \omega )}{2\omega }=D\tan ^{-1}\left(\frac{\cosh (\pi  \theta )}{\sinh (\pi 
   b)}\right).
   \label{spinonenergy}
\end{equation}

To obtain a cutoff-independent, physically meaningful energy spectrum, we perform a renormalization procedure on the spinon energy Eq.\eqref{spinonenergy}.
The minimum of the energy occurs at $\theta = 0$, defining the physical mass
\begin{equation}
    m := E_0 = D \tan^{-1}\left(\frac{1}{\sinh(\pi b)}\right).
\end{equation}
In the double scaling limit $b \to \infty$, $D \to \infty$ with $m = 2 D e^{-\pi b}$ held fixed, the energy dispersion becomes
\begin{equation}
    E_\theta \to m \cosh(\pi \theta).
\end{equation}
This shows that the originally massless chiral particle in the ultraviolet (UV) regime acquires a relativistic massive dispersion in the infrared (IR), with a well-defined mass $m$ independent of the volume and UV cutoff imposed while solving the Bethe Ansatz equations. 

In order to study the excitations on top of the ground state, we shall change the quantum numbers $I_\alpha$ by adding holes in the $n$-string sea and also add $p$-strings over various lengths $p\neq n$. For example, adding $p-$strings to the $n-$strings sea, we obtain the Bethe Ansatz equations

\begin{align}
	&\sum_{\upsilon=\pm}\left(\sum_{j=1}^n N \tan ^{-1}\left( \frac{2(\Lambda_\alpha+\upsilon b)}{2j-1}\right)+\tan ^{-1}\left( \frac{2(\Lambda_\alpha+\upsilon d)}{n}\right) \right)+\tan ^{-1}\left( \frac{2\Lambda_\alpha}{n}\right)\nonumber\\
	&=\sum_{\beta+1}^M\sum_{\upsilon=\pm}\left(\tan ^{-1}\left( \frac{\Lambda_\alpha +\upsilon \Lambda_\beta}{n}\right) +2\sum_{\xi=1}^{n-1} \tan ^{-1}\left(\frac{(\Lambda_\alpha +\upsilon \Lambda_\beta)}{\xi} \right)\right)+ \pi I_\alpha\nonumber\\
    &+\sum_{\xi=1}^\infty \sum_{\beta=1}^{M_p} \sum_{\upsilon = \pm} \left[
\tan^{-1}\left(\frac{\Lambda_\alpha + \upsilon \Lambda_\beta^\xi}{|n-p|}\right)
+ 2 \sum_{j=1}^{\frac{n+p - |n-p|}{2} - 1}
\tan^{-1}\left(\frac{\Lambda_\alpha + \upsilon \Lambda_\beta^\xi}{|n-p| + 2j}\right)
+ \tan^{-1}\left(\frac{\Lambda_\alpha + \upsilon \Lambda_\beta^\xi}{n+p}\right)
\right]
\label{all-strings-eqn}
\end{align}

We repeat the analysis performed for the ground state for this state made up of $p$-strings on top of the sea of $n$-strings. Sending $\Lambda_\alpha \to \infty$ leads to the upper bound on the quantum numbers $I_\alpha$:

\begin{equation}
    I^\infty = n N + \frac{3}{2} - M (n - 1) - \sum_{p < n} 2 p M_p - \sum_{p > n} 2 n M_p,
\end{equation}

with $I^\infty - \frac{1}{2}$ slots available, the counting can be expressed as
\begin{equation}
    n N + 1 - M (2 n - 1) - \sum_{p < n} 2 p M_p - \sum_{p > n} 2 n M_p = M + h,
\end{equation}
which can be rearranged to

\begin{equation}
M = \frac{N}{2} + \frac{1 - h - \sum_{p < n} 2 p M_p}{2 n} - \sum_{p > n} M_p.
\end{equation}

The change in the Fourier-transformed density $\Delta \tilde{\rho}_n$ due to the addition of a $p$-string is given by

\begin{equation}
\Delta \tilde{\rho}_p =
\begin{cases}
    - e^{-\frac{|\omega|}{2}(n-p)} \dfrac{1 - e^{-|\omega| p}}{1 - e^{-|\omega| n}} \cos(\omega\Lambda^p), & \text{for } p < n, \\[12pt]
    - e^{-\frac{|\omega|}{2}(p-n)} \cos(\omega\Lambda^p), & \text{for } p > n,
\end{cases}
\end{equation}
where $\Lambda^p$ is the position of the center of the $p-$string.

Using Eq.\eqref{Efinal-all-strings} and Eq.\eqref{energyeqn}, we obtain the energy of $p-$string solution due to the backflow of $n-$string solution is

\begin{equation}
   \Delta E_p=\sum _{j=1}^{\min (n,p)} -\tan ^{-1}\left(\frac{2
   (b+\Lambda  \upsilon )}{2 j-n+p-1}\right).
\end{equation}

Moreover, the bare energy of these roots is
\begin{equation}
    E_{p,\rm bare}=\sum _{j=1}^{\min (n,p)} \tan ^{-1}\left(\frac{2
   (b+\Lambda  \upsilon )}{2 j-n+p-1}\right).
\end{equation}

Hence, the total energy contribution due to the $p-$strings of length $p\neq n$ vanishes in the thermodynamic limit, and all the energy contribution comes only from the holes.

After the discussion of the construction of a generic excited state, let us focus on the most elementary excitations in the model. 

\begin{itemize}
    \item Spinless charge expectations: these gapless excitations arise from changes in the charge degrees of freedom that leave the spin quantum numbers unchanged. Consider an initial quantum number $n_j^0$ satisfying
$
- K \leq \frac{2\pi}{L} n_j^0 < 0,
$
where $K > \left|\frac{2\pi}{L} n_j\right|$ acts as a cutoff defining the "bottom of the sea" in the fully interacting theory, serving as the reference point for studying excitations. If the quantum number is shifted to
$
n_j' = n_j^0 + \Delta n \geq 0,
$
then the corresponding change in energy is given by
\begin{equation}
    \Delta E_c = \frac{2\pi}{L} \Delta n > 0.
\end{equation}

\item Chargeless spin excitations: these gapped excitations are generated by altering the spin quantum numbers $I_\alpha$, which creates holes in the sea of $n$-strings.

\begin{itemize}
    \item Triplet excitation: From the counting argument above, creating two holes in the sea of $n$-string solutions and adding a one $(n-1)$-string solution leads to a solution with total number of roots $M_t=n-1+M$ where the number of total $n-$string roots $M=n(\frac{N}{2}-1)$ such that the spin of this excitation above the ground state is
    $S^z=\frac{n}{2}N-M_t=1$. This shows that it is a triplet excitation and the energy of the excitation is $E_t=m(\cosh(\pi \theta_1)+\cosh(\pi\theta_2))$ as the energy of the one $(n-1)$-string vanishes in the thermodynamic limit.

    \item Singlet excitation: Creating two holes in the sea of solutions of $ n$ strings and adding a solution of one $n-1$ string and one $n+1$ string leads to a solution with a total number of roots $M_s=n-1+n+1+M$ where the total number of roots $n-$ string roots is $M=n\left(\frac{N}{2}-2\right)$ such that the spin of this excitations above the ground state is $S^z=\frac{n}{2}N-M_s=0$. This shows that it is a singlet excitation with energy $E_s=m(\cosh(\pi \theta_1)+\cosh(\pi \theta_2))$ where $\theta_1$ and $\theta_2$ are the positions of the two holes and as shown above the energy of $n-1$ string solution and $n+1$ string solution vanishes in the thermodynamic limit. 
\end{itemize}
All other excitations are constructed by adding even numbers of holes on top of ground state configuration made up of all $n-$string solutions and appropriate numbers of string solutions of various lengths $p\neq n$ such that the energy of the state depends only on the number of holes i.e for $2j$ numbers of holes, the degenerate excitations have energy $E_{2j}=\sum_{k}^{2j} m \cosh(\pi \theta_k)$.
\end{itemize}

Although our analysis is carried out for real values of $d$, we observe that for purely imaginary values $d = i \delta$ with $0 < \delta < \frac{1}{2}$, all results remain valid through a straightforward analytic continuation. In the overscreened Kondo phase, all excitations are bulk excitations consisting of holes and bulk string solutions. In contrast, the other phases feature additional boundary excitations layered on top of these bulk excitations.

\subsection{The zero-mode phase}
When the RG-invariant parameter $\delta$ lies within the parametric range $\frac{1}{2}<\delta<\frac{n}{2}$, the impurity is overscreened by a multiparticle Kondo cloud, similar to the overscreened Kondo phase discussed above. However, unlike the overscreened Kondo phase, boundary excitations can also arise in this regime.

The Bethe Ansatz equations for the $n-$strings sea now take the form

\begin{align}
    &\prod_{\upsilon = \pm 1} \prod_{j=1}^{n} \left(\frac{\Lambda_{\alpha} + \upsilon b + \frac{i}{2}(2j - 1)}{\Lambda_{\alpha} + \upsilon b - \frac{i}{2}(2j - 1)}\right)^N
    \frac{\Lambda_{\alpha} + i \frac{n}{2}}{\Lambda_{\alpha} - i \frac{n}{2}} 
    \prod_{\upsilon = \pm 1} \frac{\Lambda_{\alpha} +i\left(\frac{ n}{2}+\upsilon \delta\right)}{\Lambda_{\alpha}  -i\left(\frac{ n}{2}+\upsilon \delta\right)} \\
    &= - \prod_{\beta=1}^{M} \prod_{\upsilon = \pm 1} \frac{\Lambda_{\alpha} + \upsilon \Lambda_{\beta} + i n}{\Lambda_{\alpha} + \upsilon \Lambda_{\beta} - i n}
    \prod_{\xi=1}^{n - 1} \left(\frac{\Lambda_{\alpha} + \upsilon \Lambda_{\beta} + i \xi}{\Lambda_{\alpha} + \upsilon \Lambda_{\beta} - i \xi}\right)^2.
    \label{sbaenZM}
\end{align}

Following the same routine as before, we express the Bethe Ansatz equations in logarithmic form, then consider the equations for $\Lambda_{\alpha+1}$ and $\Lambda_\alpha$. Subtracting these and expanding to leading order in the difference $\Delta \Lambda_\alpha$, we obtain the  integral equation
\begin{equation}
    2 \rho_n(\Lambda) = g(\Lambda)- \sum_{\upsilon=\pm}\int \mathrm d\Lambda' K_n(\Lambda + \upsilon\Lambda')  \rho_n(\Lambda') + \mathcal{O}\left(\frac{1}{N}\right),
    \label{zmrootdens}
\end{equation}
with
\begin{align}
    g(\Lambda) &= \sum_{\upsilon=\pm}\left(\sum_{j=1}^n N a_{\frac{2j-1}{2}}(\Lambda+\upsilon b)+a_{\frac{n}{2}+\upsilon \delta}(\Lambda)\right)+a_{\frac{n}{2}}(\Lambda)-\delta(\Lambda)\\    K_n(\Lambda)&=a_n(\Lambda)+2\sum_{\xi=1}^{n-1} a_{\xi}(\Lambda
    );\qquad a_\gamma(\Lambda) = \frac{1}{\pi} \frac{\gamma}{\Lambda^2 + \gamma^2},
\end{align}
where $\delta(\Lambda)$ is added to remove the trivial root $\Lambda=0$ which leads to non-normalizable eigenfunction. Now, we add a single hole at position $\theta$ to make the total number of roots an integer as discussed above. The root density expressed in Eq.\eqref{zmrootdens} can be efficiently solved in Fourier space to obtain the solution of the form

\begin{align}
    \tilde{\rho}_n(\omega)&=\frac{e^{\frac{n | \omega | }{2}} \left(2N \cos (2 b) e^{-\frac{1}{2} (n-1) | \omega | } \left(e^{n | \omega | }-1\right)+\left(e^{|
   \omega | }-1\right) \left(-e^{\frac{n | \omega | }{2}} (2 \cos (\theta  \omega )+1)+2 \cosh (\delta \omega )+1\right)\right)}{2\left(e^{|
   \omega | }+1\right) \left(e^{n | \omega | }-1\right)}.
\end{align}
This root density is just the analytic continuation of the root density in the ground state of the overscreened Kondo phase obtained in 

Such that the total number of roots in the ground state is $M=n\int\mathrm{d}\Lambda\rho_n(\Lambda)=n\frac{N}{2}$, such that the total spin in the ground state is 
\begin{equation}
    S^z=\frac{nN+1}{2}-M=\frac{1}{2}.
\end{equation}

Just as in the overscreened Kondo phase, the impurity is overscreened by the multiparticle cloud of gapped quasiparticles in this phase. All the bulk excitations discussed above are also valid excitations in this phase. However, as shown below, apart from the bulk excitations, the model also hosts boundary excitations in this phase.

On top of the bulk string solutions, there exists a unique boundary string solution of the Bethe Ansatz equation Eq.\eqref{spineqn} of the form
\begin{equation}
    \lambda_\delta=\pm i\left(\frac{1}{2}-\delta\right),
\end{equation}
such that the Bethe Ansatz equation can be written as 
\begin{align}
&\left( \frac{\lambda_\alpha - i \left( \delta-\frac{1}{2} \right)}{\lambda_\alpha +i \left(\delta- \frac{1}{2} \right)} \right)\left( \frac{\lambda_\alpha - i \left( \frac{3}{2}-\delta \right)}{\lambda_\alpha 
+i \left(\frac{3}{2}-\delta\right)} \right)\prod_{\upsilon = \pm} \left( \frac{\lambda_\alpha + \nu b + i \frac{n}{2}}{\lambda_\alpha + \nu b - i \frac{n}{2}} \right)^N 
 = \prod_{\upsilon = \pm} \prod_{\beta \neq \alpha} \frac{\lambda_\alpha + \upsilon \lambda_\beta + i}{\lambda_\alpha + \upsilon \lambda_\beta - i}.
\label{zmbaeBSadd}
\end{align}

Bethe Ansatz equations written for all $n-$string solutions then become
\begin{align}
&\sum_{\upsilon=\pm}\left(\sum_{j=1}^n N \tan ^{-1}\left( \frac{2(\Lambda_\alpha+\upsilon b)}{2j-1}\right)\right)-\tan ^{-1}\left( \frac{2\Lambda_\alpha}{2+n-2\delta}\right) -\tan ^{-1}\left( \frac{2\Lambda}{n-2+2\delta}\right) +\tan ^{-1}\left( \frac{2\Lambda_\alpha}{n}\right)\nonumber\\
	&=\sum_{\beta=1}^M\sum_{\upsilon=\pm}\left(\tan ^{-1}\left( \frac{\Lambda_\alpha +\upsilon \Lambda_\beta}{n}\right) +2\sum_{\xi=1}^{n-1} \tan ^{-1}\left(\frac{(\Lambda_\alpha +\upsilon \Lambda_\beta)}{\xi} \right)\right)+ \pi I_\alpha,
\end{align}

Once again, we obtain the integral equation for the density of roots as
\begin{equation}
    2 \rho_n(\Lambda) = h(\Lambda)- \sum_{\upsilon=\pm}\int \mathrm d\Lambda' K_n(\Lambda + \upsilon\Lambda')  \rho_n(\Lambda') + \mathcal{O}\left(\frac{1}{N}\right),
    \label{YSRrootdens}
\end{equation}
with
\begin{align}
    h(\Lambda) &= \sum_{\upsilon=\pm}\left(\sum_{j=1}^n N a_{\frac{2j-1}{2}}(\Lambda+\upsilon b)\right)+a_{\frac{n-2+2\delta}{2}}(\Lambda)+a_{\frac{n+2-2\delta}{2}}(\Lambda)+a_{\frac{n}{2}}(\Lambda)-\delta(\Lambda)\\    K_n(\Lambda)&=a_n(\Lambda)+2\sum_{\xi=1}^{n-1} a_{\xi}(\Lambda
    );\qquad a_\gamma(\Lambda) = \frac{1}{\pi} \frac{\gamma}{\Lambda^2 + \gamma^2},
\end{align}
where $\delta(\Lambda)$ is added to remove the trivial root $\Lambda=0$ which leads to non-normalizable eigenfunction. Moreover, we add a $n-1$ string solution with center at $\bar\Lambda$  and a hole at position $\theta$ such that the solution is immediate in the Fourier space
\begin{align}
     \tilde{\rho}_{n,\delta,\bar\lambda}(\omega)&=\frac{2N \cos(2b)\sum_{j=1}^n e^{-\frac{2j-1}{2}|\omega|}-e^{-\frac{n-2+2\delta}{2}|\omega|}-e^{-\frac{n+2-2\delta}{2}|\omega|}+e^{-\frac{n}{2}|\omega|}-1}{2(1+e^{-n|\omega|}+\sum_{\xi}^{n-1}2 e^{-\xi|\omega|} )}\nonumber\\
     &-\frac{\cos(\omega\bar\Lambda)\left(e^{-\frac{|\omega|}{2}} + 2 \sum_{k=2}^{n-1} e^{-\frac{|\omega|}{2}(2k-1)} + e^{-\frac{|\omega|}{2}(2n -1)}.\right)+\cos(\omega\theta)}{(1+e^{-n|\omega|}+\sum_{\xi}^{n-1}2 e^{-\xi|\omega|})},
\end{align}
such that the total number of roots is $M_{\delta,\bar\Lambda}=1+n-1+M$ where $1$ is the number of boundary string roots, $n-1$ is the number of roots from $n-1$ strings and the number of all $n-$string solutions is $M=n\int\rho_{n,\delta,\bar\lambda}(\lambda)\mathrm{d}\lambda=n\left(\frac{N}{2}-1\right)$ and hence the total magnetization of the state is $S^z=\frac{nN}{2}+\frac{1}{2}-M_{\delta,\bar\Lambda}=\frac{1}{2}$ which is the spin of the freely propagating spinon in the bulk. 

As we have already shown, the energy of $p-$string solutions with $p\neq n$ vanishes in the thermodynamic limit. We shall now compute the energy of the added boundary string solution. The energy of the boundary string due to the shift in the root of $n-$strings sea is obtained using Eq.\eqref{Efinal-all-strings} and Eq.\eqref{energyeqn} as
\begin{align}
    \Delta E_{\delta}=-D \tan ^{-1}\left(\frac{b}{-\delta
   +\frac{n}{2}+\frac{1}{2}}\right)-D \tan
   ^{-1}\left(\frac{b}{\delta
   +\frac{n}{2}-\frac{1}{2}}\right)
\end{align}

The bare energy of the boundary string root is
\begin{align}
    E_{\delta,\rm bare}&=D \tan ^{-1}\left(\frac{2 \left(b-i
   \left(\frac{1}{2}-\delta \right)\right)}{n}\right)+D
   \tan ^{-1}\left(\frac{2 \left(b+i
   \left(\frac{1}{2}-\delta \right)\right)}{n}\right)\nonumber\\
   &=D \tan ^{-1}\left(\frac{b}{-\delta
   +\frac{n}{2}+\frac{1}{2}}\right)+D \tan
   ^{-1}\left(\frac{b}{\delta
   +\frac{n}{2}-\frac{1}{2}}\right)
\end{align}

Thus, the total energy of the boundary string solution $\Delta E_\delta + E_{\delta,\rm bare}$ vanishes in the thermodynamic limit. 

Starting from a ground state composed of all $n$-string solutions along with a single hole, we construct an excitation by adding one $(n-1)$-string and a boundary string solution. In the thermodynamic limit, this excitation carries zero energy.

By introducing a suitable number of $p$-string solutions and holes, boundary string solutions can be combined to form valid excitations. For instance, adding the boundary string solutions and $2n - 2$ holes to the ground state enables the construction of such excitations. The root density of such an excitation is
\begin{align}
     \tilde{\rho}_{n,\delta,\theta_i}(\omega)&=\frac{2N \cos(2b)\sum_{j=1}^n e^{-\frac{2j-1}{2}|\omega|}-e^{-\frac{n-2+2\delta}{2}|\omega|}-e^{-\frac{n+2-2\delta}{2}|\omega|}+e^{-\frac{n}{2}|\omega|}-1}{2(1+e^{-n|\omega|}+\sum_{\xi}^{n-1}2 e^{-\xi|\omega|} )}\nonumber\\
     &-\frac{\sum_{i=1}^{2n-2}\cos(\omega\theta_i)+\cos(\omega\theta)}{(1+e^{-n|\omega|}+\sum_{\xi}^{n-1}2 e^{-\xi|\omega|})},
\end{align}
where parameters $\theta_i$ denote the positions of the $2n - 2$ added holes, while $\theta$ represents the position of the hole in the ground state. The total number of roots in this configuration is given by

\begin{equation}
1 + n \int \rho_{n,\delta,\theta_i}(\lambda)    d\lambda = \frac{1}{2}(N - 2) n + 1,
\end{equation}

corresponding to a spin value $S^z = n - 2$. The energy of this excitation is large and expressed as

\begin{equation}
E_{\theta_i} = m \sum_{\theta_i} \cosh(\pi \theta_i),
\end{equation}
which takes the minimum value $\min(E_{\theta_i})=(n-2)m$ when each $\theta_i\to\infty$.

\subsection{The YSR Phase}
When the RG-invariant parameter $\delta$ takes a value in the parametric range $\frac{n}{2}<\delta<\frac{n}{2}+1$, then the sub-gap mode constructed from the boundary string solution gets a finite mass $E_\delta$ less than the mass gap $2m$. In this regime, the Bethe Ansatz equations for the $n-$strings sea become
\begin{align}
    &\prod_{\upsilon = \pm 1} \prod_{j=1}^{n} \left(\frac{\Lambda_{\alpha} + \upsilon b + \frac{i}{2}(2j - 1)}{\Lambda_{\alpha} + \upsilon b - \frac{i}{2}(2j - 1)}\right)^N
    \frac{\Lambda_{\alpha} + i \frac{n}{2}}{\Lambda_{\alpha} - i \frac{n}{2}} 
     \frac{\Lambda_{\alpha} +i\left(\frac{ n}{2}+ \delta\right)}{\Lambda_{\alpha}  -i\left(\frac{ n}{2}+ \delta\right)}
     \frac{\Lambda_{\alpha} -i\left(\delta-\frac{ n}{2} \right)}{\Lambda_{\alpha}  +i\left( \delta-\frac{ n}{2}\right)}
     \\
    &= - \prod_{\beta=1}^{M} \prod_{\upsilon = \pm 1} \frac{\Lambda_{\alpha} + \upsilon \Lambda_{\beta} + i n}{\Lambda_{\alpha} + \upsilon \Lambda_{\beta} - i n}
    \prod_{\xi=1}^{n - 1} \left(\frac{\Lambda_{\alpha} + \upsilon \Lambda_{\beta} + i \xi}{\Lambda_{\alpha} + \upsilon \Lambda_{\beta} - i \xi}\right)^2.
    \label{sbaenYSR}
\end{align}

Then, following the same procedure, we obtain the root density in the Fourier space as
\begin{equation}
    \tilde{\rho}_n(\omega)=\frac{2N \cos(2b)\sum_{j=1}^n e^{-\frac{2j-1}{2}|\omega|}-2 e^{-\delta  | \omega | } \sinh \left(\frac{n | \omega |
   }{2}\right)+e^{-\frac{n}{2}|\omega|}-1}{2(1+e^{-n|\omega|}+\sum_{\xi}^{n-1}2 e^{-\xi|\omega|} )},
\end{equation}

such that the total number of roots in the ground state is $M=\frac{n N}{2}$ and the spin in this state is $\frac{nN}{2}+\frac{1}{2}-M=\frac{1}{2}$. This state contains no holes and is formed only from $n$-string solutions; hence, its spin-$\frac{1}{2}$ comes entirely from the unscreened impurity.

However, it is possible to construct a state where the impurity is screened by a sub-gap bound mode which is localized at the edge. To construct such a state, we shall add a single hole at position $\theta$, the boundary string solution $\lambda_\delta$, and an $(n-1)$-string with center $\bar\Lambda$ such that the root density becomes
\begin{align}
    \tilde{\rho}_n(\omega)&=\frac{2N \cos(2b)\sum_{j=1}^n e^{-\frac{2j-1}{2}|\omega|}+e^{-\frac{n}{2}|\omega|}-1-2\cos(\omega\theta)-e^{-\frac{n-2+2\delta}{2}|\omega|}-e^{-\frac{n+2-2\delta}{2}|\omega|}}{2(1+e^{-n|\omega|}+\sum_{\xi}^{n-1}2 e^{-\xi|\omega|} )},\nonumber\\
    &-\frac{\cos(\omega\bar\Lambda)\left(e^{-\frac{|\omega|}{2}} + 2 \sum_{k=2}^{n-1} e^{-\frac{|\omega|}{2}(2k-1)} + e^{-\frac{|\omega|}{2}(2n -1)}.\right)}{(1+e^{-n|\omega|}+\sum_{\xi}^{n-1}2 e^{-\xi|\omega|})}.
\end{align}

The total number of roots is $M_{\delta,\bar\Lambda}=1+n-1+M$ where $1$ is the number of boundary string roots, $n-1$ is the number of roots from $n-1$ strings and the number of all $n-$string solutions is $M=n\int\rho_{n,\delta,\bar\lambda}(\lambda)\mathrm{d}\lambda=n\left(\frac{N}{2}-1\right)$ and hence the total magnetization of the state is $S^z=\frac{nN}{2}+\frac{1}{2}-M_{\delta,\bar\Lambda}=\frac{1}{2}$ which is the spin of the freely propagating spinon in the bulk while the impurity is screened by this localized bound mode.

As before, the energy of the $(n-1)$-string solution vanishes in the thermodynamic limit, and the energy of the single hole at position $\theta$ is $E_\theta=m\cosh(\pi \theta),$ which takes the minimum value $\min(E_\theta)=m$ when $\theta\to\infty$. We now compute the energy of the boundary string solution by first noting that the change in the root density of the $n$-string sea due to the presence of the boundary string solution is 
\begin{equation}
    \Delta\rho_{n}(\omega)=\frac{-e^{-\frac{n-2+2\delta}{2}|\omega|}-e^{-\frac{n+2-2\delta}{2}|\omega|}+2\sinh\left(|\omega| \frac{n}{2}\right)e^{-\delta|\omega|}}{2(1+e^{-n|\omega|}+\sum_{\xi}^{n-1}2 e^{-\xi|\omega|} )}.
\end{equation}
The energy contribution of the boundary string solution due to the shift in the density of the $n-$ string solutions obtained by using Eq.\eqref{Efinal-all-strings} and Eq.\eqref{energyeqn} is
\begin{equation}
    \Delta E_{\delta}=-D \tan ^{-1}\left(\frac{b}{-\delta
   +\frac{n}{2}+\frac{1}{2}}\right)-D \tan
   ^{-1}\left(\frac{b}{\delta
   +\frac{n}{2}-\frac{1}{2}}\right)-D \arctan\left(\frac{\cos\left(\frac{\pi}{2}(n - 2\delta)\right)}{\sinh(\pi b)}\right),
\end{equation}
such that adding the bare energy of the boundary string solutions cancels the first two terms, and taking the scaling limit $D\to\infty$ and $b\to\infty$, we obtain the total energy of the bound mode 
\begin{equation}
    E_\delta= -m \cos\left(\frac{\pi}{2}(n - 2\delta)\right).
\end{equation}
Hence, the energy of this excitation on top of the ground state, which is composed entirely of $n$-string solutions with the impurity remaining unscreened, is obtained by adding a single hole, the boundary string solution, and an $(n-1)$-string solution. The total energy is
\begin{equation}
E = m \left(1 - \cos\left(\frac{\pi}{2}(n - 2\delta)\right)\right),
\end{equation}
where the first term corresponds to the minimum energy of the spinon, the second term arises from the boundary string solution, and the energy contribution from the $(n-1)$-string solutions vanishes in the thermodynamic limit. In this state, the impurity is screened by this single-particle bound mode.

Thus, the ground state undergoes a drastic change at $\delta = \frac{n}{2}$. For $\delta < \frac{n}{2}$, the impurity is screened in the ground state, whereas for $\delta > \frac{n}{2}$ it remains unscreened. However, in the narrow parametric window
$
\frac{n}{2} < \delta < \frac{n}{2} + 1,
$
the impurity is screened by a mid-gap excitation. As we will see, once $\delta > \frac{n}{2} + 1$, the mid-gap state ceases to exist, and the impurity cannot be completely screened in any eigenstate.

\subsection{Unscreened Phase}

As the RG-invariant parameter $\delta$ further increases and exceeds $\frac{n}{2} + 1$, the superconducting order in the bulk dominates over the Kondo coupling strength at the boundary, preventing the impurity from being fully screened in this phase. 

The ground state in this phase consists of $n$-strings and can be obtained as the analytic continuation of the ground state in the YSR phase to the parametric regime $\delta > \frac{n}{2} + 1$, where the impurity remains unscreened.

The boundary string solution $\lambda_\delta$ is still a valid solution in this regime. Adding this solution on top of the ground state results in a change in the root density of the $n-$string sea by 
\begin{equation}
    \Delta\rho_{n}(\omega)=\frac{-e^{-\frac{n-2+2\delta}{2}|\omega|}+e^{-\frac{2\delta-n-2}{2}|\omega|}+2\sinh\left(|\omega| \frac{n}{2}\right)e^{-\delta|\omega|}}{2(1+e^{-n|\omega|}+\sum_{\xi}^{n-1}2 e^{-\xi|\omega|} )},
\end{equation}

such that the energy due to the backflow of the $n-$strings in the presence of this root obtained from  Eq.\eqref{Efinal-all-strings} and Eq.\eqref{energyeqn} is exactly canceled by the bare energy of the root, such that $E_\delta=0$. As this root cannot be added without including an appropriate even number of holes and various bulk strings, the resulting state lies above the mass gap; more importantly, the impurity remains unscreened in this state.

\section{Derivation of renormalization group equations}\label{RGderivation-Nch}
We obtained the expression for the superconducting mass gap
\begin{equation}
m=D \arctan\left(\frac{1}{\sinh(\pi b)}\right).
\end{equation}
Upon taking the scaling limit $D\to \infty$ and $b\to \infty$ while holding $m$ fixed, we write
\begin{equation}
    m=2De^{-\pi b}\approx 2De^{-\frac{\pi}{2g}}.
\end{equation}
Inverting this relation, we obtain
\begin{equation}
    \frac{1}{2b(D)}=g(D)=\frac{\pi }{2 \ln \left(\frac{2 D}{m}\right)}.
    \label{gflow}
\end{equation}
Moreover, from the expression of the RG-invariant quantity $d$, we obtain
\begin{equation}
    c=\frac {2 b} {b^2 - d^2 - 1},
\end{equation}
such that for small $J$ by taking $c=2J$  we obtain
\begin{equation}
    J(D)=\frac { b} {b^2 - d^2 - 1} .
    \label{jflow}
\end{equation}

Using Eq.~\eqref{gflow} in Eq.~\eqref{jflow} and differentiating Eq.~\eqref{gflow} and Eq.~\eqref{jflow} with respect to $\ln D$, we arrive at the RG equations
\begin{align}
    \beta(g)=\frac{\mathrm{d}}{\mathrm{d}\ln D}g&=-\frac{2}{\pi}g^2,\\
    \beta(J)=\frac{\mathrm{d}}{\mathrm{d}\ln D}J&=-\frac{2}{\pi } J (J-g).
\end{align}

These RG equations govern the RG flow in our cut-off scheme, which coincides with more conventional schemes - lattice of momentum cut-off - at low energy. Therefore, the conclusions derived from the RG equations are reliable only for weak coupling. The complete understanding of the phase diagram requires a full solution of the model.

\section{Thermodynamics in overscreened Kondo phase}\label{TBA-OSK}
Under the string hypothesis, one assumes that all solutions of  the Bethe Ansatz equations are string solutions of the form 
\begin{equation}
    \lambda_j^{(p)} = \lambda^{(p)} + i\frac{1}{2}(p + 1 - 2j), \quad j = 1, 2, \ldots, p.
\end{equation}

Writing the Bethe Ansatz equations for all $p-$strings and taking $\ln$ on both sides of the equation, we obtain
\begin{equation}
  \Theta_{p}(\lambda^{(p)}_\gamma)+\sum_\nu  \Theta_{p}(\lambda^{(p)}_\gamma+\nu  d)+N\sum_{j=1}^{\min(p,n)} \Theta_{n+p+1-2j}(\lambda^{(p)}_\gamma+\nu  b)=\sum_{m,\beta}\Theta_{p,m}(\lambda^{(p)}_\gamma+\nu \lambda^{(m)}_\beta)-2\pi I_\gamma^{(p)},
\end{equation}

where
\begin{equation}
    \Theta_p(x)=-2 \tan^{-1}\left(\frac{2 x}{p}\right),
\end{equation}

and
\begin{equation}
    \Theta_{p m}(x) = \begin{cases} 
\Theta_{|p-m|}(x) + 2\Theta_{|p-m|+2}(x) + \cdots + 2\Theta_{p+m-2}(x) + \Theta_{p+m}(x), & p \neq m \\ 
2\Theta_2(x) + \cdots + 2\Theta_{2p-2}(x) + \Theta_{2p}(x), & p = m .
\end{cases}
\end{equation}

The counting function 
\begin{equation}
    \begin{aligned}
        \nu_p(\lambda) = \frac{1}{2\pi}\bigg[&\Theta_{p}(\lambda^{(p)}_\gamma)+\sum_\nu  \Theta_{p}(\lambda^{(p)}_\gamma+\nu  d)+N\sum_{j=1}^{\min(p,n)} \Theta_{n+p+1-2j}(\lambda^{(p)}_\gamma+\nu  b)-\sum_{m,\beta}\Theta_{p,m}(\lambda^{(p)}_\gamma+\nu \lambda^{(m)}_\beta) \bigg]
    \end{aligned}
\end{equation}
is such that it gives the integers $I^{(p)}_\gamma$ for corresponding roots $\lambda^{(p)}_\gamma$, i.e.\ $\nu_p(\lambda^{(p)}_\gamma)=I^{(p)}_\gamma$, and the skipped integers $I^{(p),h}_\gamma$ correspond to the positions of holes, i.e.\ $\nu_p(\lambda^{(p),h}_\gamma)=I^{(p),h}_\gamma$.          

The derivative of the counting function in the thermodynamic limit gives the density of $p$-strings $\rho_p(\mu)$ and holes $\rho_p^h(\mu)$:
\begin{equation}
    \frac{\mathrm d \nu_p}{\mathrm d \mu} = \rho_p(\mu) + \rho_p^h(\mu).
\end{equation}

Combining the last two expressions relates the density of holes to that of the density of the string solutions via
\begin{equation}\label{sigma_p^h}
    \rho_p^h(\mu) = f_p(\mu) - \sum_{m=1}^{\infty}A_{p m} \rho_m(\mu),
\end{equation}

where
\begin{equation}
    \begin{aligned}
        f_p(\mu) &= a_{\frac{p}{2}}(\mu)+\sum_{\upsilon=\pm}\left(N\sum_{j=1}^{\min(p,n)} a_{\frac{n+p+1-2j}{2}}(\mu +\upsilon b) + a_{\frac{p}{2}}(\mu +\upsilon d) \right), \\
        A_{p m} &= \left[ |p - m| \right] + 2 \left[ |p - m| + 2 \right] + \cdots + 2 \left[ p + m - 2 \right] + \left[ p + m \right],
    \end{aligned}
\end{equation}

with $a_{\frac{p}{2}}(\mu)$ defined as
\begin{equation}
    a_{\frac{p}{2}}(\mu) \equiv -\frac{1}{2\pi} \frac{\mathrm d \Theta_p}{\mathrm d \mu} = \frac{1}{\pi} \frac{\frac{p}{2}}{\left( \frac{p}{2} \right)^2 + \mu^2},
\end{equation}

and the functional $\left[ p \right]$ introduced as convolution with $a_{\frac{p}{2}}$:
\begin{equation}
    \left[ p \right] g(\mu) \equiv a_{\frac{p}{2}} \star g(\mu) = \int d\lambda    a_{\frac{p}{2}}(\mu - \lambda) g(\lambda).
\end{equation}

Moreover, the energy function becomes
\begin{equation}
    E=\frac{2\pi}{L} \sum_{j=1}^{N^e} n_j+D\sum_{j=1}^{\min(p,n)}\int\mathrm{d}\lambda \rho_p(\lambda) \Theta_{n+p+1-2j}(\upsilon\lambda+  b)
\end{equation}

In Eq.\eqref{Efinal-all-strings}, we note that the first term is the contribution from the charge energy \textit{i.e.}
the charge sector energy is given by
\begin{equation}
E^{(c)}(\{n_j\}) = \frac{2\pi}{L} \sum_{j=1}^{N^e} n_j,
\end{equation}
where $n_j$ are the quantum numbers associated with the charge degrees of freedom. The corresponding charge partition function is
\begin{equation}
Z^{(c)} = \sum_{\{n_j\},   n_j \geq -N^e} \exp\left(-\frac{1}{T} \sum_{j=1}^{N^e} \frac{2\pi}{L} n_j \right).
\end{equation}
This partition function characterizes a system of $N^e$ noninteracting spinless fermions possessing a linear dispersion relation. When the parameter $D$ becomes very large, the free energy simplifies to
\begin{equation}
F^{(c)} = -\frac{L T}{2 \pi} \int_{-\infty}^\infty dk    \ln \left(1 + e^{-\frac{k}{T}}\right) = -\frac{\pi}{12} L T^2 + \{\text{divergent constant}\}.
\end{equation}

The spin sector’s free energy, when subjected to a magnetic field $H$, can be expressed as
\begin{equation}
    \mathcal{F}=E+2M H - T\mathcal{S},
\end{equation}
where $\mathcal{S}$ denotes the Yang-Yang entropy, which can be approximated using Stirling’s formula as

\begin{equation}
    \mathcal{S}=\sum_{p=1}^\infty \int \mathrm{d} \lambda   \left[ (\rho_p+\rho_p^h)\ln(\rho_p+\rho_p^h)-\rho_p\ln\rho_p - \rho_p^h\ln\rho_p^h \right].
\end{equation}
The combination $E+M H$ can thus be written as
\begin{equation}
    E+M H =\sum_{p=1}^\infty \int \mathrm{d} \lambda    g_p(\lambda) \rho_p(\lambda),
\end{equation} 
and introducing $$g_p(\lambda)=2pH-D\left(\sum_{j=1}^{\min(p,n)}\Theta_{n+p+1-2j}(\upsilon\lambda+  b)-2n\pi\right),$$ one can write the free energy as \footnote{Note that the free boundary also contributes to the free energy through a ratio of determinants of two Gaudin matrices. However, our focus is on the impurity contribution, which is obtained by taking the difference in thermodynamic entropy between an open chain with and without the impurity, where these free boundary contributions cancel out.}
\begin{equation}\label{F}
    \mathcal{F}=\sum_{p=1}^\infty \int \mathrm{d} \lambda    \left[ g_p \rho_p - T\rho_p \ln\left[1+\frac{\rho_p^h}{\rho_p}\right] - T\rho_p^h \ln\left[1+\frac{\rho_p}{\rho_p^h}\right] \right].
\end{equation}
Varying the free energy subjected to the constraint $\delta \rho_p^h = -\sum_{m=1}^\infty A_{pm}\delta \rho_m$  from Eq.\eqref{sigma_p^h} we get
\begin{equation}
    g_p-T\ln\left[1+\frac{\rho_p^h}{\rho_p}\right]+T\sum_{m=1}^\infty A_{pm}\ln\left[1+\frac{\rho_m}{\rho_m^h}\right]=0.
\end{equation}
or, introducing $\eta_p = \rho_p^h/\rho_p$, one can write
\begin{equation}\label{pre TBA}
\ln\left[1+\eta_p(\lambda)\right]=\frac{g_p(\lambda)}{T}+\sum_{m=1}^\infty A_{pm} \ln\left[1+\eta^{-1}_m(\lambda)\right]  .
\end{equation}

It is convenient to introduce a functional $G$ acting by convolution with $1/2\cosh(\pi \lambda)$
\begin{equation}
    Gf(\lambda) = \int \mathrm{d} \mu ~ \frac{1}{2\cosh\pi (\lambda-\mu)}f(\mu).
\end{equation}

Applying $\delta_{m,p}-G(\delta_{m-1,p}+\delta_{m+1,p})$ to the Eq.\eqref{pre TBA}, we obtain
\begin{equation}\label{TBA}
    \ln \eta_p(\lambda) = 
   -\frac{m}{T}\cosh(\pi \lambda)\delta_{n,p}+G\ln\left[ 1+\eta_{p+1} \right]+G\ln\left[ 1+\eta_{p-1} \right].
\end{equation}

Applying $G$ on Eq.\eqref{pre TBA}, we obtain
\begin{align}
    G\left[\ln[1+\eta_p(\lambda)]-\frac{g_p(\lambda)}{T} \right]=\sum_{m=1}^\infty Y_{p,m}\ln[1+\eta_p^{-1}(\lambda)]
\end{align}
where we introduced 
    \begin{equation}
        Y_{p,m}(\mu)=\sum_{l=1}^{\mathrm{min}(p,m)} a_{\frac{p+m+1-2l}{2}}(\mu).
\end{equation}
Noticing
\begin{equation}
    Y_{1,m}(\mu)=a_\frac{m}{2}(\mu),
\end{equation}
we can simplify the equation for free energy as 
\begin{equation}
    \begin{aligned}
        \mathcal{F}&=\frac{1}{2}\int \mathrm{d} \lambda   \left\{ \left( \frac{N}{2\cosh \pi (\lambda-b)}+\frac{N}{2\cosh \pi (\lambda+b)}\right)\left[ g_{n}(\lambda)-T\ln(1+\eta_{n}(\lambda)) \right]\right.\\
        &\left.+\left( \frac{1}{2\cosh\pi \lambda}+\frac{1}{2\cosh\pi (\lambda-d)}+\frac{1}{2\cosh\pi (\lambda+d)} \right)\left[g_1(\lambda)-T\ln(1+\eta_1(\lambda))\right] \right\}.
    \end{aligned}
\end{equation}
The impurity part of the free energy is
\begin{equation}
    \begin{aligned}
        \mathcal{F}_{\mathrm{imp}}&=\mathcal{F}_{\mathrm{imp}}^0-\frac{T}{2}\int \mathrm{d} \lambda  \left(\frac{1}{2\cosh\pi (\lambda-d)}+\frac{1}{2\cosh\pi (\lambda+d)} \right)\ln(1+\eta_1(\lambda)).
    \end{aligned}
    \label{freeenegg}
\end{equation}
These are the equations studied in the main text.

We refer the reader to Ref.~\cite{kattel2025thermodynamics} for a detailed discussion of the thermodynamics in the remaining phases featuring multiple towers. In the Appendix of that work, we explicitly demonstrate how summing over the tower contributions yields the free energy presented in the main text.

\section{Numerical solution of TBA equations}\label{sec:NTBA}
We solve the infinite hierarchy of thermodynamic Bethe Ansatz (TBA) equations for the impurity entropy by truncating the string index at a finite cutoff $n_{\max}$. The TBA equations

\begin{equation}
\ln \eta_p(\lambda) = -\frac{m}{T} \cosh\left(\pi \lambda\right) \delta_{p, n} + \sum_{\upsilon = \pm 1} \int_{-\infty}^\infty \frac{d\mu}{2 \cosh\left[\pi \left(\lambda - \mu\right)\right]} \ln \left[ 1 + \eta_{p+\upsilon}(\mu) \right]
\end{equation}

are solved numerically on a discrete rapidity grid $\lambda_i \in \left[-L, L\right]$, with $L=L_{\rm max}$. The convolution kernel is discretized as

\begin{equation}
G_{ij} = \frac{\Delta \lambda}{2 \cosh\left[\pi \left(\lambda_i - \lambda_j\right)\right]},
\end{equation}

transforming the integral equations into matrix-vector form. The hierarchy is truncated by imposing the large-$p$ boundary condition at $p = n_{\max} + 1$,

\begin{equation}
\ln \left( 1 + \eta_{n_{\max}+1}(\lambda) \right)== \frac{n_{\max} + 1}{n_{\max}} \ln \left( 1 + \eta_{n_{\max}}(\lambda) \right),
\end{equation}

allowing closure of the system. We iteratively update $\eta_p(\lambda_i)$ until convergence.

The impurity contribution to the free energy is computed as

\begin{equation}
F_{\mathrm{imp}}(T) = - \frac{T}{2} \sum_{\upsilon = \pm 1} \int_{-\infty}^\infty \frac{d\lambda}{2 \cosh\left[\pi \left(\lambda + \upsilon d\right)\right]} \ln \left[ 1 + \eta_1(\lambda) \right],
\end{equation}

approximated by a discrete sum over the grid points. The impurity entropy follows from numerical differentiation,

\begin{equation}
S_{\mathrm{imp}}(T) = - \frac{d F_{\mathrm{imp}}}{d T}.
\end{equation}

\begin{figure}[H]
  \centering
  \subfigure[Impurity entropy $S_{\mathrm{imp}}$ for the six-channel Kondo model as a function of the universal scaling variable $T/m$, obtained by numerically solving the Thermodynamic Bethe Ansatz (TBA) equations. Different colors represent distinct values of the RG-invariant parameter $d$ (blue: $d=1$, red: $d=0.5$, green: $d=0$). Curves with markers ($+$) are for $m=0.5$ and dashed lines are for $m=1$. The overlay of these curves explicitly demonstrates the universal collapse of data as a function of $T/m$. The entropy smoothly interpolates between the low-temperature limit $S_{\mathrm{imp}}(T \to 0) = \ln\left(\frac{1}{2}(1+\sqrt{5}\right)$ and the high-temperature limit $S_{\mathrm{imp}}(T \to \infty) = \ln 2$, illustrating universal scaling behavior $S_{\mathrm{imp}} = S_{\mathrm{imp}}(d, T/m)$.]{
    \includegraphics[width=0.48\textwidth]{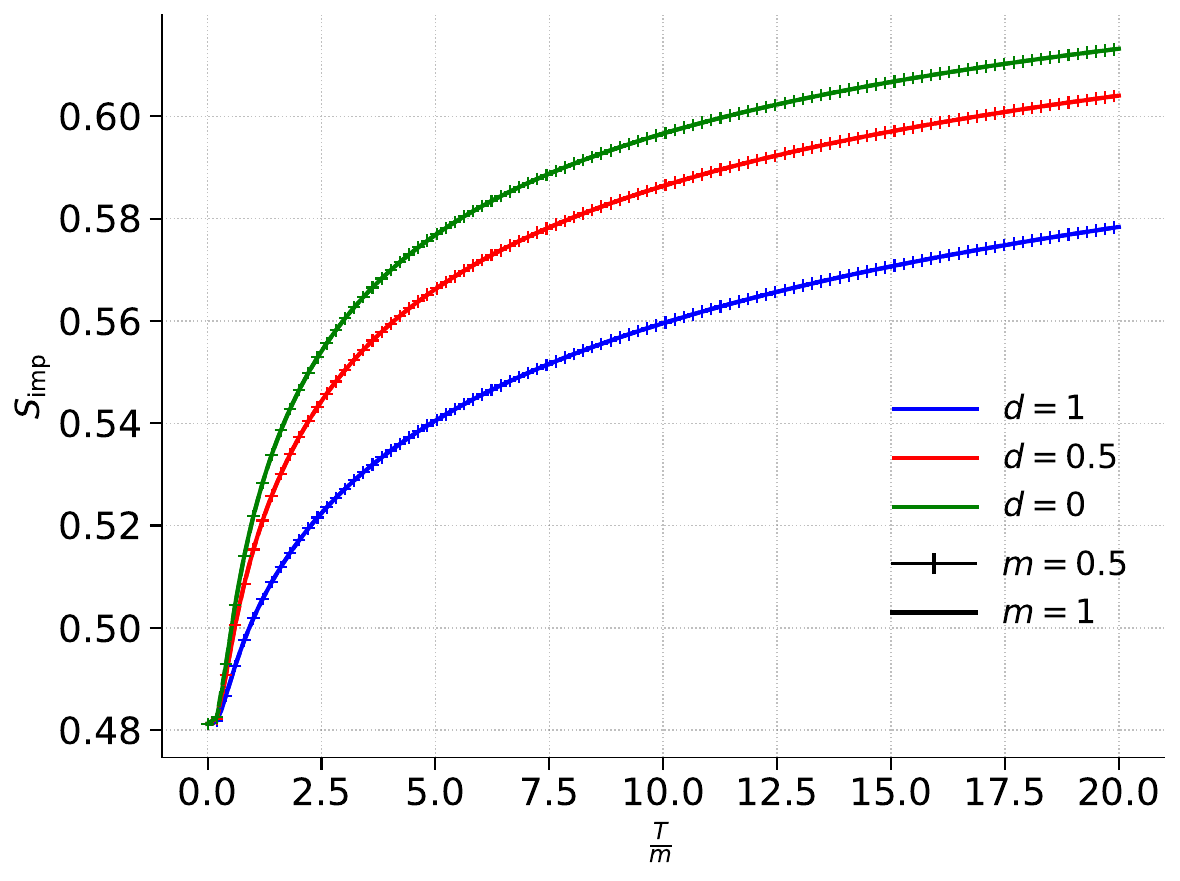}
    \label{fig:3-channel-fig}
  }
  \hfill
  \subfigure[Impurity entropy $S_{\mathrm{imp}}$ for the conventional multichannel Kondo model obtained in the limit $g=0$, plotted as a function of the universal scaling variable $T/T_0$, with the Kondo scale defined as $T_0 = D e^{-\pi/c}$. Different curves correspond to varying channel numbers $n = 1$ through $6$, demonstrating well-known universal crossover behavior. This figure serves as a benchmark for comparison: our model, incorporating bulk interactions and a finite mass gap, reproduces exactly the same infrared (IR) and ultraviolet (UV) fixed-point entropies shown here. However, the crossover trajectories in our model explicitly depend on an additional RG-invariant parameter $d$.
]{
    \includegraphics[width=0.48\textwidth]{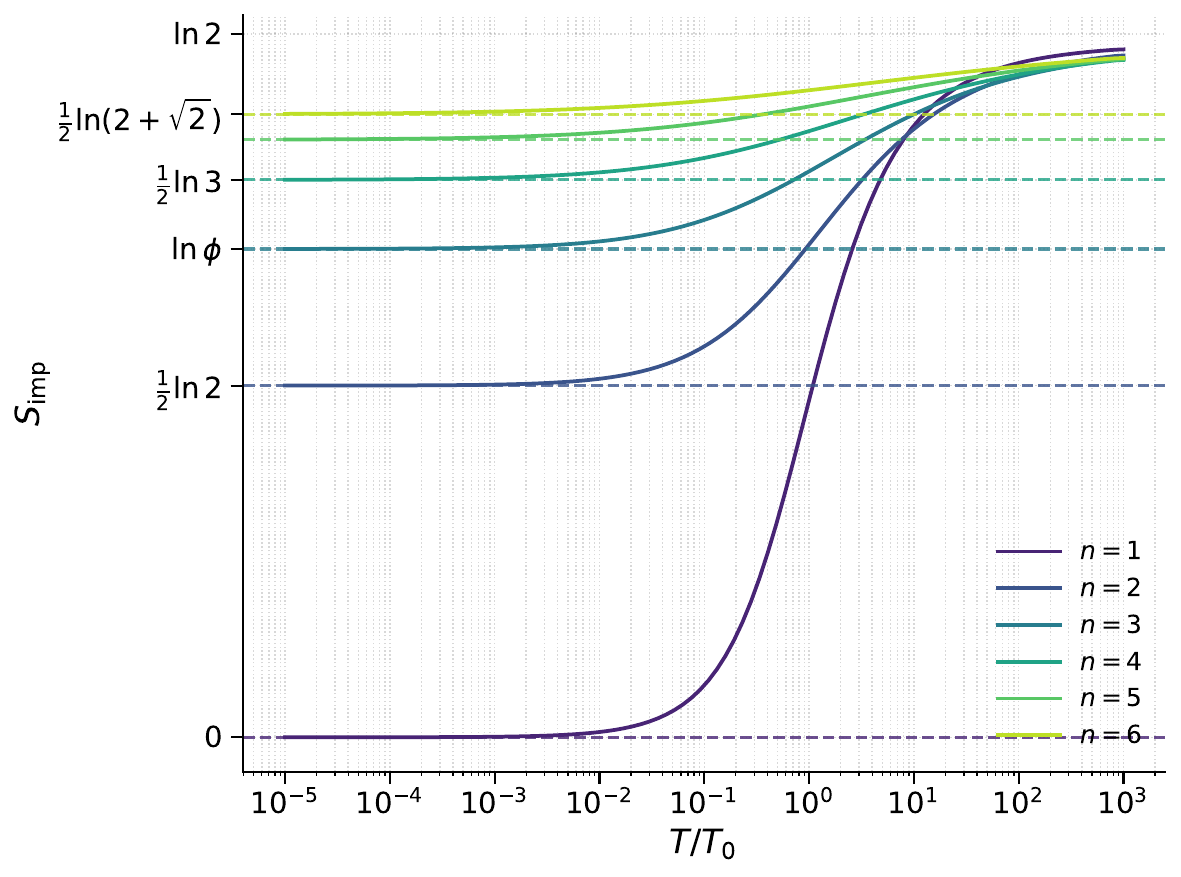}
    \label{fig:multichannel-tba}
  }
  \caption{Impurity entropy from numerical solution of TBA a) for $g\neq 0$ and b) $g=0$}
  \label{fig:entropy-comparison}
\end{figure}

We now turn to solving the TBA equations numerically for n=3. Notice that the convolution kernel $G(\lambda)$ is sharply peaked at zero and decays exponentially away from the origin. This property ensures excellent numerical convergence even with moderate rapidity cutoffs. Using $L_{\max} = 20$ for the rapidity range and truncating the string hierarchy at $n_{\max} = 20$, we numerically solve the TBA equations to compute the impurity entropy as shown in Fig.\ref{fig:3-channel-fig}. We explore parameters $d = 0,   0.5,   1$ and $m = 0.5,   1$. At low temperatures, the impurity entropy approaches
\begin{equation}
S_{\mathrm{imp}}\left(T \to 0\right) \approx 0.48 \approx \ln \phi,
\end{equation}
where $\phi = \frac{1 + \sqrt{5}}{2}$ is the golden ratio, consistent with the known residual entropy of the three-channel Kondo fixed point. 

In Fig.\ref{fig:multichannel-tba}, we obtained the impurity entropy in the $g=0$ limit when the model under consideration reduces to the conventional Kondo problem, which was extensively studied in Refs.~\cite{andrei1984solution,tsvelick1984solution}.

\section{Low temperature asymptotics and critical exponents}\label{sec:lowtemp-asymptotics}
Defining a new variable $ \zeta = \pi \lambda $, so that $ d\lambda = d\zeta / \pi $, the free energy Eq.\eqref{freeenegg} can be written as
\begin{equation}
F_{\rm{imp}}(T) = -\frac{T}{2\pi} \int_{-\infty}^\infty d\zeta    e^{-\frac{m}{T} \cosh \zeta}
\left[ \frac{1}{\cosh(\zeta + \pi d)} + \frac{1}{\cosh(\zeta - \pi d)} \right].
\end{equation}

Moreover, defining the kernel
\begin{equation}
K_d(\zeta) = \frac{1}{\cosh(\zeta + \pi d)} + \frac{1}{\cosh(\zeta - \pi d)},
\end{equation}
the expression can be compactly written as
\begin{equation}
F_{\rm{imp}}(T) = -\frac{T}{2\pi} \int_{-\infty}^\infty d\zeta    e^{-\frac{m}{T} \cosh \zeta} \cdot K_d(\zeta).
\end{equation}

At low $T$, the driving term makes $\eta_k(\lambda)$ exponentially small away from $\lambda = 0$, and $\eta_1(\lambda)$ inherits a similar decay
\begin{equation}
    \ln(1 + \eta_1(\lambda)) \sim \exp\left(-\frac{m}{T} \cosh(\pi \lambda)\right).
\end{equation}

The kernel consists of two terms centered at $\lambda = \pm d$. The integrand is the product of a sharp peak at $\lambda = 0$ and two broader peaks at $\lambda = \pm d$. The dominant contribution to the integral arises from the overlap between the exponential tails of $\ln(1 + \eta_1(\lambda))$ and the kernel peaks, which occurs at $\lambda>>1$.

The exponential suppression of the driving term at the kernel peak $ \zeta =\pm \pi d $ is $
\exp\left( -\frac{m}{T} \cosh(\pi d) \right)$.
Therefore, the integrand becomes $ \mathcal{O}(1) $ when $\frac{m}{T} \cosh(\pi d) \sim 1.$ This defines the Kondo temperature
\begin{equation}
{T_K = m \cosh(\pi d)}.
\end{equation}

To derive the subleading power-law correction to the impurity free energy at low temperature, we analyze the linearized form of the TBA equations. 
We define
\begin{equation}
g_p(\zeta) = \ln(1 + \eta_p(\zeta)), \qquad \delta g_p(\zeta) = g_p(\zeta) - g_p(\infty).
\end{equation}

At low temperature, the corrections $\delta g_p$ are small, and the TBA equations linearize. For $p < n$, we have
\begin{equation}
\frac{f_p^2}{f_{p+1} f_{p-1}} \delta g_p(\zeta) = G * \left( \delta g_{p-1}(\zeta) + \delta g_{p+1}(\zeta) \right),
\end{equation}
where $f_p^2 = 1 + \eta_p(\infty)$, $f_0 = 1$, and $\delta g_n(\zeta) = g_n(\zeta)$ is the driving term.

Taking the Fourier transform gives a set of algebraic recursions for $\delta \tilde{g}_p(\omega)$. These can be solved in terms of a transfer function
\begin{equation}
\delta \tilde{g}_p(\omega) = \hat{t}_{p,n}(\omega) \cdot \tilde{g}_n(\omega),
\end{equation}
with
\begin{equation}
\hat{t}_{p,n}(\omega) = \frac{f_{p-1} \sinh[\omega(p+2)/2] - f_{p+1} \sinh[\omega p/2]}{2 f_p \cos[\pi/(n + 2)] \sinh[(n/2 + 1) \omega]}.
\end{equation}

The impurity free energy correction becomes
\begin{equation}
\begin{aligned}
\mathcal{F}_{\mathrm{imp}}(T) &= \mathcal{F}_{\mathrm{imp}}^0 - T \ln\left( \frac{\sin[2\pi/(n + 2)]}{\sin[\pi/(n + 2)]} \right) \\
&\quad - T \int \frac{d\omega}{2\pi}    e^{-i \omega \ln(T/2\pi)/\pi} \cdot \frac{\cos(\omega d)}{2 \cosh(\omega/2)} \cdot \hat{t}_{1,n}(\omega) \cdot \tilde{g}_k(\omega).
\end{aligned}
\end{equation}

The integrand has poles at
\begin{equation}
\omega = \frac{2\pi i m}{n + 2}, \quad m \in \mathbb{N},
\end{equation}
arising from the $\sinh[(n/2 + 1)\omega]$ in the denominator. The first pole at $\omega = \frac{\pi i}{n/2 + 1}$ has vanishing residue, so the next pole at
\begin{equation}
\omega_* = \frac{2\pi i}{n + 2}
\end{equation}
dominates.

Evaluating the residue yields the leading non-analytic correction
\begin{equation}
\delta \mathcal{F}_{\rm{imp}}(T) \sim T^{1 + \frac{4}{n + 2}}.
\end{equation}

Thus, the specific heat behaves as
\begin{equation}
{C_{\rm{imp}}(T) \sim T^{\frac{4}{n + 2}}},
\end{equation}
and the critical exponent is
\begin{equation}
{\alpha = - \frac{4}{n + 2}}.
\end{equation}

In the special case $n = 2$, the second pole becomes a double pole, and the free energy correction becomes logarithmic:
\begin{equation}
C_{\rm{imp}}(T) \sim T \ln(1/T).
\end{equation}

We now compute the magnetic field exponent $\delta$ by analyzing the zero-temperature impurity magnetization in the presence of a small magnetic field $H \ll T_K$.

We begin with the ground state root density contribution to the impurity in the Fourier space
\begin{equation}
\rho_n^{\mathrm{imp}}(\omega) =
\frac{e^{\frac{n|\omega|}{2}} (e^{|\omega|} - 1)}{(e^{|\omega|} + 1)(e^{n|\omega|} - 1)} \cdot \cos(d\omega).
\end{equation}

The singularities of this function occur at
\begin{itemize}
  \item $\omega = i\pi(2m + 1), ~ m\in \mathbb{Z}$ from $e^{|\omega|} + 1 = 0$,
  \item $\omega = \frac{2\pi i m}{n}, ~ m \in \mathbb{Z}$ from $e^{n|\omega|} - 1 = 0$.
\end{itemize}
The closest pole to the real axis is:
\begin{equation}
{\omega_* = \frac{2\pi i}{n}}.
\end{equation}

This determines the large-$\lambda$ decay of the real-space impurity root density
\begin{equation}
\rho_n^{\mathrm{imp}}(\lambda) \sim e^{-2\pi \lambda / n}, \quad \lambda \to \infty.
\end{equation}

Now, at zero temperature, the applied magnetic field introduces a Fermi cutoff at rapidity $\lambda_H$, defined by the condition
\begin{equation}
\cosh(\pi \lambda_H)\sim \frac{1}{2}e^{\pi \lambda_H} \sim \frac{m}{H} \quad \Rightarrow \quad
\lambda_H \sim \frac{1}{\pi} \ln\left( \frac{2m}{H} \right).
\end{equation}

The impurity magnetization is then given by the integral over the unfilled tail
\begin{equation}
M(H) = \int_{\lambda_H}^\infty \rho_n^{\mathrm{imp}}(\lambda)    d\lambda
\sim \int_{\lambda_H}^\infty e^{-2\pi \lambda / n} d\lambda
= \frac{n}{2\pi} e^{-2\pi \lambda_H / n}.
\end{equation}

Substituting the expression for $\lambda_H$, we obtain
\begin{equation}
M(H) \sim \left( \frac{2m}{H} \right)^{-2/n} \propto H^{2/n}.
\end{equation}

Hence, the magnetic exponent is
\begin{equation}
{\delta = \frac{n}{2}}.
\end{equation}

In the limit $d \gg 1$, the kernel $K_d(\zeta) = \frac{1}{\cosh(\zeta + \pi d)} + \frac{1}{\cosh(\zeta - \pi d)}$ consists of two peaks sharply localized around $\zeta = \pm \pi d$. Each contributes independently to the free energy, and the integrand $e^{- \frac{m}{T} \cosh \zeta}$ is dominated near these peaks.

Focusing near $\zeta = \pi d$, set $x := \zeta - \pi d$. Using $\cosh(x + \pi d) \approx \frac{1}{2} e^{\pi d + x}$ for $d \gg 1$, we find $e^{- \frac{m}{T} \cosh(\zeta)} \approx \exp\left( - \frac{m}{2T} e^{\pi d + x} \right)$. Define the Kondo scale $T_K := \frac{m}{2} e^{\pi d}$ so that this becomes $\exp(- \frac{T_K}{T} e^x)$. Similarly, expanding near $\zeta = -\pi d$ yields a symmetric contribution $\exp(- \frac{T_K}{T} e^{-x})$.

Combining both gives:
\begin{equation}
F_{\rm imp}(T) = -\frac{T}{2\pi} \int_{-\infty}^\infty dx \left[ e^{- \frac{T_K}{T} e^x} + e^{- \frac{T_K}{T} e^{-x}} \right] \frac{1}{\cosh x}.
\end{equation}
Defining $t := T/T_K$, the impurity free energy becomes:
\begin{equation}
F_{\rm imp}(T) = T \cdot f(t), \quad f(t) := -\frac{1}{2\pi} \int_{-\infty}^\infty dx \left[ e^{-e^x/t} + e^{-e^{-x}/t} \right] \frac{1}{\cosh x}.
\end{equation}

This scaling function $f(t)$ captures the universal crossover in the $d \gg 1$ regime and depends only on $T/T_K$ as mentioned in the main text.

\subsection{Some remarks on Boundary Infrared Fixed Point and Conformal field theory}

The low-temperature behavior of the multichannel Kondo model with a gapless bulk is governed by a nontrivial infrared fixed point described by an $SU(2)_n$ Wess–Zumino–Witten (WZW) conformal field theory. The spin sector is described by the affine algebra $\widehat{\mathfrak{su}}(2)_n$, whose integrable highest-weight representations arise as quotients of Verma modules by null submodules. These exist only for  
$j = 0, \frac{1}{2}, 1, \ldots, \frac{n}{2}$,  
yielding exactly $n + 1$ primary fields. The truncation ensures unitarity and modular covariance, leading to a finite modular $S$-matrix

\begin{equation}
S_{ji} = \sqrt{\frac{2}{n + 2}} \sin\left( \frac{(2j + 1)(2i + 1)\pi}{n + 2} \right).
\end{equation}

In the Affleck–Ludwig boundary CFT framework, each consistent conformal boundary condition is labeled by one of these primaries. The impurity contribution to the zero-temperature entropy is given by the logarithm of the quantum dimension

\begin{equation}
S_{\mathrm{imp}} = \ln d_j,
\qquad
d_j = \frac{S_{0j}}{S_{00}} = \frac{\sin\left( \frac{(2j + 1)\pi}{n + 2} \right)}{\sin\left( \frac{\pi}{n + 2} \right)}.
\end{equation}

The quantity $d_j$ measures the effective degeneracy of the boundary state and is generally fractional for $n \ge 2$. Its noninteger value reflects the nontrivial fusion algebra and the fractionalized impurity degrees of freedom at the infrared fixed point. The same conformal data determine the scaling dimension of the spin operator

\begin{equation}
\Delta = \frac{2}{n + 2},
\end{equation}

which governs the boundary spin correlations. Thus, the universal low-energy physics—including the residual entropy and the critical exponents—is fully determined by the $SU(2)_n$ WZW conformal field theory describing the infrared fixed point.

In the thermodynamic Bethe Ansatz (TBA) formulation, the same quantity $d_j$ emerges directly from the exact solution of the microscopic Hamiltonian. Starting from the Bethe Ansatz equations, one canonically diagonalizes the many-body Hamiltonian by introducing rapidities $\{\lambda_\alpha\}$ satisfying coupled algebraic equations determined by the scattering phase shifts. Each admissible set of rapidities corresponds to an eigenstate, and the full partition function is obtained by summing over all allowed Bethe configurations.

In the scaling limit, the system is described by integral TBA equations for the pseudoenergies $\epsilon_m(\lambda)$ of the string species $m = 1, \ldots, n$. The impurity contribution to the free energy is extracted as a boundary correction to the bulk thermodynamic potential,

\begin{equation}
F_{\mathrm{imp}}(T) = -T \sum_{m=1}^{n} \int d\lambda\, a_m(\lambda) \ln\left( 1 + e^{-\epsilon_m(\lambda)/T} \right),
\end{equation}

where $a_m(\lambda)$ are known kernels derived from the Bethe scattering matrix. At low temperature, the $\epsilon_m(\lambda)$ satisfy coupled integral equations whose infrared fixed point encodes the residual degeneracy of the screened impurity.

Evaluating the free energy in the limit $T \to 0$ gives the impurity entropy

\begin{equation}
S_{\mathrm{imp}} = -\frac{\partial F_{\mathrm{imp}}}{\partial T}\bigg|_{T\to 0}
= \ln d_j,
\end{equation}

with the same $d_j$ that appears in the $SU(2)_n$ WZW theory. Specifically, in the UV, the impurity behaves as a free spin of dimension $2S + 1$, while in the IR, the bound-state structure implied by the fusion hierarchy yields an effective degeneracy proportional to

\begin{equation}
d_j = \frac{\sin\left( \frac{(2j + 1)\pi}{n + 2} \right)}{\sin\left( \frac{\pi}{n + 2} \right)}.
\end{equation}

The Bethe Ansatz solution first revealed the presence of this universal number through the exact canonical diagonalization of the Hamiltonian and the explicit summation over all eigenstates to construct the full partition function. In the thermodynamic limit, the analysis of the TBA equations showed that the impurity entropy saturates at a noninteger residual value, indicating an overscreened fixed point. It was only later that Affleck and Ludwig provided the conformal field theory interpretation, identifying this number with the logarithm of the quantum dimension in the $SU(2)_n$ WZW model. Their work demonstrated that the critical point uncovered microscopically by the Bethe Ansatz corresponds precisely to an $SU(2)_n$ conformal fixed point, with the same fractional degeneracy and scaling exponents predicted by the WZW conformal data.

In the present case, the situation is more subtle: the bulk is no longer conformal due to the presence of a finite mass gap, yet the impurity entropy remains equal to $\ln d_j$ over a wide range of parameters and low temperatures. This persistence of the fractional boundary entropy indicates that, even though bulk conformal invariance is broken, the impurity degrees of freedom continue to flow to an effective $SU(2)_n$ boundary fixed point for a large parametric regime. As the parameters are further tuned, a quantum phase transition occurs beyond which the impurity is no longer overscreened and instead retains its full $\ln 2$ entropy at low temperatures, signaling a return to a decoupled, local-moment regime.

As shown in the main text, the Bethe Ansatz uncovers this transition microscopically through a change in the structure of the excitation towers: new boundary scales emerge, and some of the low-lying towers are lifted in energy. This reorganization of the Bethe spectrum encodes the crossover from the fractionalized to the fully localized impurity state.

It would be interesting to reexamine this behavior from the perspective of perturbative boundary conformal field theory. When the bulk coupling $g = 0$, the infrared criticality is well understood in terms of the $SU(2)_n$ WZW model, as discussed above. However, for $g \neq 0$, conformal symmetry is broken in the bulk in the IR due to the opening of the gap in the spin sector, and the full spectrum is no longer conformal—correlation functions acquire finite correlation lengths and the continuous tower of conformal descendants is replaced by discrete gapped excitations. Nevertheless, because the impurity entropy continues to take the value $\ln d_j$ across an extended regime, some remnant of the boundary conformal structure may persist. Understanding how this residual structure survives in the presence of a nonconformal bulk, and how it can be captured as a deformation of the boundary CFT by relevant bulk operators, would be an interesting direction for further analysis.

\end{document}